\documentclass[sigconf]{acmart}
\usepackage[utf8]{inputenc}
\usepackage[T1]{fontenc}
\usepackage{textcomp}
\usepackage{xcolor,colortbl}
\usepackage{multirow}
\usepackage{soul}
\usepackage{hyperref} 
\usepackage{array}
\usepackage{graphicx} 
\usepackage{subcaption}
\usepackage{booktabs}
\usepackage{tabularx}
\usepackage{makecell}

\AtBeginDocument{%
  \providecommand\BibTeX{{%
    \normalfont B\kern-0.5em{\scshape i\kern-0.25em b}\kern-0.8em\TeX}}}

\newif\ifdraft
\drafttrue

\newcommand{\modi}[1]{\textcolor{black}{#1}}

\copyrightyear{2026}
\acmYear{2026}
\setcopyright{cc}
\setcctype{by-nc-nd}
\acmConference[CHI '26]{Proceedings of the 2026 CHI Conference on Human Factors in Computing Systems}{April 13--17, 2026}{Barcelona, Spain}
\acmBooktitle{Proceedings of the 2026 CHI Conference on Human Factors in Computing Systems (CHI '26), April 13--17, 2026, Barcelona, Spain}
\acmPrice{}
\acmDOI{10.1145/3772318.3791003}
\acmISBN{979-8-4007-2278-3/2026/04}

\begin{document}


\begin{teaserfigure}
  \centering
  \includegraphics[width=\textwidth]{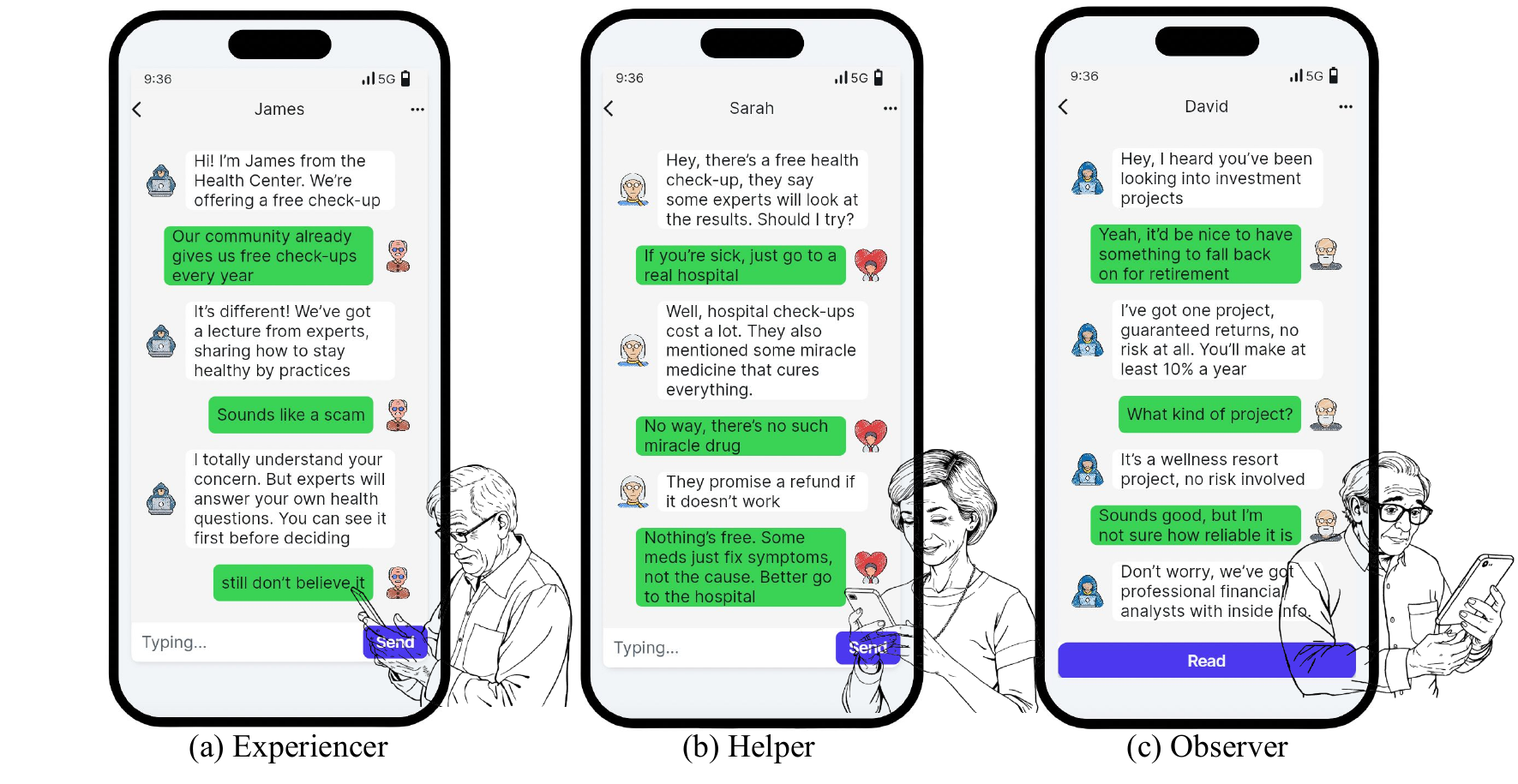}
  \caption{Three role-based simulation perspectives for online fraud intervention in a learning activity when older adults (a) experience simulated online fraud; (b) help a victim of online fraud; or (c) observe online fraud cases.}
  \Description{Figure 1: Three smartphone interface mock-ups illustrate role-based perspectives for online fraud intervention among older adults. (a) Experiencer: An older adult directly engages in a suspicious conversation, learning through personal experience. (b) Helper: An older adult advises another person who is at risk of fraud, learning by providing support. (c) Observer: An older adult passively reviews a fraud conversation, learning by observation.}
  \label{illustration}
\end{teaserfigure}

\title[Online Fraud Intervention for Older Adults]{Experiencer, Helper, or Observer: Online Fraud Intervention for Older Adults Through a Role-based Simulation Approach}

\author{Yue Deng}
\affiliation{%
  \institution{Hong Kong University of Science and Technology}
  \city{Hong Kong}
  \country{China}
}
\affiliation{
  \institution{Max Planck Institute for Security and Privacy}
  \city{Bochum}
  \country{Germany}
}
\email{ydengbi@connect.ust.hk}

\author{Xiaowei Chen}
\affiliation{%
  \institution{Max Planck Institute for Security and Privacy}
  \city{Bochum}
  \country{Germany}
}
\email{xiaowei.chen@mpi-sp.org}

\author{Junxiang Liao}
\affiliation{%
  \institution{Hong Kong Polytechnic University}
  \city{Hong Kong}
  \country{China}
}
\email{junxiang.liao@connect.polyu.hk}

\author{Bo Li}
\affiliation{%
 \institution{Hong Kong University of Science and Technology}
  \city{Hong Kong}
  \country{China}
}
\email{bli@cse.ust.hk}

\author{Yixin Zou}
\affiliation{%
  \institution{Max Planck Institute for Security and Privacy}
  \city{Bochum}
  \country{Germany}
}
\email{yixin.zou@mpi-sp.org}

\renewcommand{\shortauthors}{Deng et al.}

\begin{abstract}
Online fraud is a critical global threat that disproportionately targets older adults. Prior anti-fraud education for older adults has largely relied on static, traditional instruction that limits engagement and real-world transfer, whereas role-based simulation offers realistic yet low-risk opportunities for practice. Moreover, most interventions situate learners as victims, overlooking that fraud encounters often involve multiple roles, such as bystanders who witness scams and helpers who support victims. To address this gap, we developed \modi{ROLESafe}, an anti-fraud educational intervention in which older adults learn through different learning roles, including Experiencer (experiencing fraud), Helper (assisting a victim), and Observer (witnessing fraud). In a between-subjects study with 144 older adults in China, we found that the Experiencer and Helper roles significantly improved participants' ability to identify online fraud. These findings highlight the promise of role-based, multi-perspective simulations for enhancing fraud awareness among older adults and provide design implications for future anti-fraud education.
\end{abstract}



\maketitle

\section{Introduction}

Online fraud has emerged as one of the most prevalent and costly digital threats worldwide \cite{button2014online, cross2022meeting}. Older adults are disproportionately targeted due to multiple age-related risk factors \cite{judges2017role,alves2008effects,james2014correlates}. 
Their vulnerability is further exacerbated by the continuous evolution and increasing sophistication of scam tactics \cite{deng2025auntie,zhai2025hear} and AI's application in enhancing the effectiveness, speed, and credibility of fraudulent schemes \cite{herrera2024bridging}. Consequently, older adults often express growing apprehension about the complexity of such threats and the difficulty of recognizing fraud in practice \cite{sheil2024enhancing}. Fraud targeting older adults has increasingly become a global threat. In China, survey data from the Chinese Academy of Social Sciences revealed that 17.25\% of older users had experienced online fraud in 2022 \cite{url3}. According to the U.S. 
Federal Trade Commission, American victims in their 60s reported the highest financial losses to fraud, totaling \$1.18 billion in 2024 \cite{url1}. 

Moreover, online fraud is not only a technological or economic phenomenon but is also shaped by cultural and regional contexts \cite{parti2025cross,houtti2024survey}. For instance, a comparative study across 12 countries reveals variations in scam exposure, victimization, types, vectors, and reporting \cite{houtti2024survey}. In China, where the online fraud environment differs from Western contexts \cite{lee2021online}, scholars have noted that scammers often exploit local cultural narratives, such as deference to authority and collective values \cite{li2025cultural}. China is one of the world’s most populous countries and is experiencing rapid population aging \cite{lobanov2023growing}, yet online fraud targeting older adults remains understudied. We therefore situate our study in the Chinese context.

Despite numerous studies in HCI highlighting the unique vulnerabilities of older adults to online fraud and calling for more tailored interventions \cite{deng2025auntie,dai2025envisioning,oliveira2017dissecting,lin2019susceptibility}, empirical fraud intervention for this population remains scarce. Existing approaches to anti-fraud education for older adults largely rely on traditional one-way formats such as training sessions or broadcast media \cite{button2024disseminating}, which provide limited interactivity and practical exercises. More engaging approaches, such as board games \cite{chung2023reducing, camilleri2025cybersafety}, have primarily focused on phishing or general cybersecurity rather than online fraud. 

Among various educational formats, role-based simulation has been found as effective in engaging participants with learning and creating realistic learning scenarios
\cite{lain2022phishing,karumbaiah2016phishing,yeoh2022simulated,gopavaram2021cross}. ``What.Hack,'' for example, engages players in a role-playing phishing simulation where they assume the role of a bank employee tasked with processing business-related emails while avoiding phishing attempts \cite{wen2019hack}. However, existing simulations have rarely been tailored to older adults, and most primarily place participants in the role of the victims. Nevertheless, fraud usually involves complex social interaction, including not only scammers and victims, but also bystanders who observe and helpers who provide support. Building on established learning theories, different modes of participation may foster effective learning in distinct ways. For instance, Experiential Learning Theory emphasizes that knowledge is constructed through concrete experiences \cite{kolb2014experiential}. The Learning by Teaching approach underscores how explaining and guiding others can deepen one's own understanding \cite{bargh1980cognitive}. Social Learning Theory highlights that learning is situated within a social context and can occur through observation \cite{bandura1977social}. In online fraud education, different roles may thus provide unique perspectives and lead to varied learning outcomes. However, such role-based comparisons have rarely been systematically examined.
To address this gap, we propose the following research questions:

\begin{itemize}
\item \textbf{RQ1}: How does role-based simulation learning influence older adults' fraud awareness between the three treatment groups (Experiencer, Helper, Observer) and the control group (static learning material)?

\item \textbf{RQ2}: How do different roles (Experiencer, Helper, Observer) within simulation learning influence older adults' online fraud awareness? 
\end{itemize}


To investigate our research questions, we designed three simulation roles, including \textit{Experiencer}, where older adults engaged in online fraud conversations with an LLM playing the scammer, \textit{Helper}, where they persuaded an LLM-portrayed victim not to fall into fraud, and \textit{Observer}, where they watched LLM-generated chat records based on real-world fraud cues. We conducted a between-subjects experiment with 144 Chinese older adults and evaluated our intervention's effectiveness in raising online fraud awareness and user experience, while gathering participants' qualitative feedback. Our findings show that role-based simulation was more effective than the control condition in improving fraud cue identification. Both Experiencers and Helpers outperformed the Control group, and Helpers further outperformed Observers. \modi{However, these group differences were not observed in a follow-up evaluation.} Qualitative feedback informed our reimagination of fraud education design for older adults. Our work makes the following contributions to the human-computer interaction (HCI) and security and privacy (S\&P) literature.
\begin{itemize}

\item 
We develop and evaluate an online fraud intervention tailored for older adults using role-based simulation, providing learning opportunities in realistic scam scenarios to enhance online fraud awareness.

\item 
We make the first attempt to educate older adults about online fraud with different roles (Experiencer, Helper, and Observer), informing future explorations of multi-perspective approaches to fraud education.

\item 
We provide empirical evidence on the effectiveness of role-based simulation for online fraud intervention, and demonstrate how different roles shape both learning outcomes and nuanced perceptions, offering new insights into the design of anti-fraud education for older adults.
\end{itemize}

\section{Related Work}

\subsection{Online Fraud and Older Adults}
Online fraud is one of the most pressing digital threats for older adults.
The FBI's Internet Crime Complaint Center documented a 14\% increase in elder fraud cases, and tech support scams were the most widely reported kind of elder fraud in 2023 \cite{url2}. Age-related changes make older adults particularly susceptible to online fraud. Prior research has identified various risk factors, such as cognitive decline \cite{judges2017role}, an overly trusting nature \cite{shao2019older}, and a lack of knowledge about fraud prevention \cite{james2014correlates}. For example, several studies indicated older adults show higher susceptibility to phishing and are more likely to click compared to younger groups \cite{greitzer2021experimental, li2020experimental,lin2019susceptibility}.

The risks are amplified by the continuous evolution of scams. Fraud schemes now take increasingly diverse forms, such as advertising of health products on live-streaming platforms \cite{deng2025auntie} and sophisticated deepfake-enabled scams \cite{zhai2025hear}. Older adults frequently express apprehension about the growing complexity of such threats, reporting that scams are becoming more difficult to recognize and voicing a desire for more information on evolving scams \cite{sheil2024enhancing}. News coverage of severe financial losses further reinforces a pervasive fear of being targeted, leaving many older adults increasingly concerned about the threat of online fraud \cite{quan2018revisiting}.

Preventing older adults from falling victim to online fraud remains challenging. Although many detection tools exist \cite{edwards2017scamming,shen2025warned,liu2023understanding,ghosh2025temper}, their accuracy is not fully guaranteed, and the diversity of scam formats limits their applicability. Family members often play safeguarding roles, but these efforts are constrained by seniors' reluctance to accept help, intentional withholding of information, and delays in family intervention \cite{deng2025auntie}. Moreover, many victims are reluctant to report scams or seek help due to stigma or lack of accessible support \cite{parti2023if}. These barriers emphasize the importance of strengthening older adults' own fraud awareness and their ability to identify deceptive tactics.

However, most prior work stops at broadly advocating for educational online fraud interventions for older adults. 
To address this gap, our study developed \modi{ROLESafe}, an online fraud intervention designed specifically for older adults to improve their online fraud awareness and defensive skills.

\subsection{Anti-Fraud Education for Older Adults}

Anti-fraud education for older adults has thus far largely relied on traditional delivery methods. A wide range of channels have been used to disseminate fraud awareness and prevention information, including formal events, presentations, training sessions, in-person or telephone advice, letters and leaflets, emails, websites, and broadcast media such as radio and television \cite{button2024disseminating}. For example, the U.S. National Council on Aging developed the Seniors Against Scams program, delivered in senior centers, which highlights popular scams targeting vulnerable older adults and offers next steps for those who experience financial fraud \cite{url4}. Similarly, the Federal Deposit Insurance Corporation and the Consumer Financial Protection Bureau launched the Money Smart for Older Adults curriculum, consisting of an instructor guide, a resource guide, and supplemental slides to support classroom-based training \cite{url5}. While these initiatives have raised awareness and provided valuable resources, their effectiveness is limited by the fact that they are largely one-directional, offering information passively without providing interactive or experiential opportunities for older adults to practice identifying and resisting fraud.

In addition to these traditional approaches, one recent work has explored innovative methods of educating older adults about fraud. Chung and Yeung evaluated an anti-scam board game specifically designed for older adults, which simulates scam scenarios on a game board and preventative measures as playing cards \cite{chung2023reducing}. Although not exclusively focused on fraud, other cybersecurity education initiatives have incorporated scam-related content. For example, Camilleri et al. developed a shedding-type card game, which included handling scams as one of several cybersafety topics, to conceptually learn and reinforce cyber hygiene practices \cite{camilleri2025cybersafety}. Similarly, Aly et al. tailored digital privacy education interventions for older adults across multiple modalities (text, videos, audio presentations, infographics, comics, interactive tutorials, and chatbots), with fraudulent online transactions identified as one of the key concerns \cite{aly2024tailoring}. 

However, these efforts mainly emphasize general preventive awareness or post-fraud practices, with very few providing fine-grained training, that is, how seniors can realize that an interaction is fraudulent and recognize the cues that signal deception.

\subsection{Fraud Cues}
\modi{Fraud identification can be understood as a judgment process. According to Brunswik’s lens model, people make judgments by perceiving and utilizing various cues 
\cite{brunswik1955representative,cooksey1996judgment}.} In phishing, the cue utilization is presumed to involve an individual's capacity to recognize features within an email that signal an attempt to deceive \cite{nasser2020role}. \modi{A number of phishing cues have been identified, such as technical cues (e.g., URL hyperlinking), visual presentation cues (e.g., no branding or logos), and message language and content cues (e.g., use of time pressure, threatening language, and too good to be true offers) \cite{molinaro2018evaluating}. Prior work in phishing education has demonstrated the importance of cue-based training \cite{moreno2017fishing,sturman2024roles}.} For instance, Anti-Phishing Phil was developed to train users to recognize phishing cues in web browsers \cite{sheng2007anti}. Another embedded training system by Kumaraguru et al. similarly aimed to teach users how to identify phishing cues in general~\cite{kumaraguru2007protecting}.


\modi{Real-world fraud in China is rarely a single-shot decision problem; it typically unfolds through chat-based interactions in which scammers strategically exploit social interactions on popular messaging platforms \cite{deng2025auntie}. For instance, scammers often add victims on WeChat (a popular social app in China) and gradually build rapport through repeated promises and performative care \cite{deng2025auntie}. In such contexts, simply asking older adults to make a broad, binary judgment about whether a case is fraudulent fails to capture the skills they actually rely on when navigating real scam interactions. In contrast, training older adults to identify specific tactics used by scammers could more directly reflect the underlying reasoning processes and offer greater ecological validity and real-world applicability. Building on these insights, in our training materials, we define \textbf{fraud cues} as heuristics that users can draw upon to determine whether an interaction might be a scam, and these cues generally map onto deceptive tactics commonly employed by scammers. In this sense, the cues in our study are more similar to the language and content cues in the phishing literature \cite{molinaro2018evaluating} and focus less on technical details or visual presentation features.}


Nevertheless, existing cue-based training efforts in cybersecurity have predominantly focused on phishing, with limited attention to online fraud more broadly, and especially to the unique challenges faced by older adults in identifying such cues.
This gap is becoming increasingly urgent as scammers now exploit AI to enhance fraud effectiveness, speed, and credibility, thereby reducing detectability of scams that target seniors \cite{herrera2024bridging}. This shift indicates that protecting seniors should no longer rely on detecting false artifacts, but instead fostering situational awareness and a deeper understanding of how scams are executed \cite{herrera2024bridging}. To address this gap, we aim to educate older adults on how online fraud unfolds and how to identify the online fraud cues in practice through a role-based simulation approach.



\subsection{Role-based Simulation Learning}
Role-based simulation learning does not have a universally agreed-upon definition, and the terms role play and simulation are often used interchangeably \cite{linser2011can}. Broadly speaking, simulation-based learning is considered a form of experiential learning in which learners are tasked with solving complex problems in controlled environments through ``real-life scenarios'' \cite{lateef2010simulation,url6}. Role plays are situations in which learners take on the role profiles of specific characters in a contrived setting \cite{wills2012role}.

Role-based simulation learning offers several advantages. First, it enables learners to absorb knowledge and practice skills in realistic yet simulated environments, thereby providing hands-on and first-person perspectives \cite{url6,url7, lateef2010simulation}. Second, it is practical in nature, as role plays and simulations are designed to authentically emulate real-world environments or challenges, giving learners opportunities to rehearse the skills they will need in practice \cite{url8}. Third, it provides a safe and supportive environment in which learners can experiment, reflect, and practice without the fear of real-world consequences or mistakes \cite{lateef2010simulation}. Finally, role-based simulations promote engagement, helping learners to learn effectively while making the learning process enjoyable \cite{shaw2018designing}.

Role-based simulation has been widely adopted across diverse educational domains. In medical and nursing training, role-players are extensively used to help students practice clinical and conversational skills before encountering real-world patients~\cite{janssens2023maternity,wang2021they,o2016observer}. In computer science, Asanka et al. propose a platform where students actively participate in a role-based learning exercise designed to simulate real-world big data projects \cite{asanka2024learning}. In business contexts, role-play simulations are employed to prepare leaders for interpersonal conflict, negotiations, and performance reviews~\cite{url7}. 

Role-based simulation has also seen widespread application in HCI and S\&P areas, such as training about system attacks \cite{cone2007video} and phishing \cite{karumbaiah2016phishing,lain2022phishing,gopavaram2021cross,kumaraguru2008lessons,yeoh2022simulated,chen2024effects}. 
For example, CyberCIEGE is a construction and management resource simulation game in which students act as decision-makers for enterprises, facing scenarios that require balancing security decisions against organizational needs \cite{thompson2014cyberciege}. PhishGuru delivered simulated phishing attacks and provided training immediately after participants fell for them \cite{kumaraguru2008lessons}.

Despite this diversity, most role-based interventions adopt a single-role perspective, positioning participants as potential victims (i.e., experiencers) and training them to recognize and resist attacks. Yet, fraud encounters are fundamentally social interactions involving multiple roles, not only scammers and their targets, but also bystanders who observe \cite{burgoon1996deceptive,brink2015reporting} and helpers who provide support \cite{deng2025auntie}. Different perspectives may lead to different learning outcomes. However, no existing work has explored online fraud education through multiple role perspectives, nor compared the effectiveness of these roles. Therefore, this study examines the effects of role-based simulation and those of different roles (i.e., observer, helper, and experiencer) on online fraud learning for older adults.

\section{ROLESafe: An Online Fraud Intervention for Older Adults}

We developed \modi{ROLESafe (Role-Oriented Learning for Enhancing Safety)}. In \modi{ROLESafe}, older adults learn through different roles within simulated fraud scenarios to strengthen their online fraud awareness.

\subsection{Role-playing Learning Modes}
\label{learning modes}

Drawn on established learning theories and prior empirical findings, we designed three role-based learning modes that reflect different perspectives on online fraud, shown in Figure \ref{illustration}. 
\begin{itemize}
    \item \textbf{Experiencer Mode:} This mode is informed by Kolb’s Experiential Learning Theory, which emphasizes that learners acquire knowledge through concrete experience \cite{kolb2014experiential}. Prior research has similarly presented specific fraud scenarios that allow users to experience the process, thereby strengthening their resistance and awareness \cite{kumaraguru2008lessons,yeoh2022simulated,kumaraguru2008lessons,gopavaram2021cross,lain2022phishing,karumbaiah2016phishing,wen2019hack}. For example, one study emulated the normal working day of an executive assistant, requiring participants to complete work-related tasks and respond to emails while simultaneously identifying and reporting phishing attempts \cite{karumbaiah2016phishing}. Building on this, we designed an Experiencer Mode in which participants directly engaged with simulated online fraud scenarios as experiencers. They interacted with a simulated scammer through a chat interface.  

    \item \textbf{Helper Mode:} This mode is inspired by the Learning by Teaching approach, which indicates the benefits of explaining and guiding others as a means of deepening one’s own understanding \cite{bargh1980cognitive}. This helper role is in line with empirical studies demonstrating that helping others can also enhance helpers’ own awareness and resistance against online fraud or cybersecurity risks \cite{deng2025auntie,murthy2021individually,menges2023caring}. For example, the AARP field experiment asked participants to take the role of advising others not to fall for fraudulent telemarketing, and this self-generated persuasion significantly reduced the participants' own subsequent susceptibility to scams \cite{anderson2003off}. Drawing on this, we designed a Helper Mode, where participants acted as helpers by attempting to support a simulated potential victim of fraud. They interacted with the simulated vulnerable individual through the chat interface and were encouraged to provide guidance to prevent the victim from being deceived.
    
    \item \textbf{Observer Mode:} 
    This mode draws on Social Learning Theory, which highlights that learning is a cognitive process situated within a social context and can occur purely through observation, even without direct practice or reinforcement \cite{bandura1977social}. Prior research has shown that observing scam cases can also prompt participants' reflection and foster awareness \cite{aly2024tailoring,zhai2025hear}. For instance, Zhai et al. designed deepfake scam scenarios presented as videos and flash reports to elicit empathy for victims and to explore how older adults perceive such scams and why they believe some victims are deceived \cite{zhai2025hear}. Building on these insights, we designed an Observer Mode in which participants acted as observers by viewing a simulated scam conversation on the chat interface. Unlike experiencers and helpers, they did not interact with either party but instead passively observed the dialogue.

    \item \textbf{Control:} Participants in the control condition did not engage in any role-based simulation. As a baseline, they only read online fraud learning materials with details shown in \modi{the Supplementary Material}. These materials were also made available to participants in the three role-based conditions.
\end{itemize}

\subsection{Design Materials}
\label{Design Materials}

\subsubsection{Online Fraud Discourses.} Although general resources on online fraud are abundant, studies that specifically organize and analyze online fraud cues targeting older adults remain scarce. Therefore, we drew on timely and widely accessible materials from media and news sources to collect online fraud discourses targeting older adults and extract fraud cue information. 

We conducted a Google search using the keywords \textit{``older adults'' AND ``online'' AND ``fraud''} in Chinese. We reviewed the first 50 search results, with each result potentially containing multiple fraud cases. We applied the following inclusion criteria:  
1) To ensure reliability and relevance for older adult learners, we only included credible sources, such as news reports, articles published by banks, and government documents (e.g., police or prosecutor’s office announcements). The majority of the search results already fell into these categories;  
2) Only cases situated in the Chinese context were retained, as fraud practices may vary across cultures; 
3) Each case had to include concrete online fraud details, such as the fraud process and discourse. Cases that only reported fraud categories or statistical data were excluded;
4) Offline fraud cases were excluded.  
After applying these criteria, we obtained 34 results containing 103 online fraud cases targeting older adults.

\subsubsection{Online Fraud Contexts.}


\modi{For online fraud targeting older adults in China, scammers gradually probe for and exploit their specific vulnerabilities \cite{deng2025auntie, url11}. The representative vulnerabilities involve financial expectations and health or well-being concerns among older adults \cite{deng2025auntie}. For example, seniors' financial expectations can lead them to fall victim to investment fraud, while their concerns for health and well-being make them vulnerable to false advertising of health products \cite{deng2025auntie}.}

\modi{Based on the prior work \cite{deng2025auntie} and our collected cases, we focused on the two prevalent vulnerabilities as the prototype contexts for our intervention: financial expectations and health concerns. In the finance context, scammers often exploit older adults' financial expectations and gradually leverage cues such as ``promise high returns with low risk'' to defraud them, for example, by persuading them to invest in fraudulent projects. In the health context, scammers usually take advantage of seniors' concerns about health and gradually draw on cues such as ``offer free services as bait'' to lead them to purchase supposed health products or services.}

\subsubsection{Online Fraud Cues \& Channels.}
\label{fraud cue}
\modi{Our analysis of collected cases revealed that although the concrete fraudulent content may differ---for example, health-related fraud could have false advertising of health products or courses---the underlying fraud cues leveraged, such as exaggerating illness to create anxiety in older adults, are largely the same. 
To extract the online fraud cues, 
the first author conducted open coding \cite{corbin2014basics} to identify potential fraud cues. Specifically, the author first read through the collected online fraud discourses multiple times and extracted all segments that contained any concrete detail about how the scam was carried out. 
Based on these extracted
segments, the first author developed initial codes that demonstrated the fraud cues. 
For example, in the segment where ``the project manager tells Mr. Qiao that a blockbuster film can earn several times the returns and carries no risk,'' we coded this as the fraud cue ``promising high returns with no risk.''
Through several rounds of team discussions, the co-authors reviewed the extracted segments and codes and worked collaboratively to refine the fraud cues. 
Finally, we consolidated a set of fraud cues, including nine cues in the finance context and seven cues in the health context.\footnote{\modi{Regarding whether cues within each context are mutually exclusive, the cues of each context in our list do not overlap with one another, but we cannot guarantee mutual exclusivity across all real-world fraud contexts.}}}

\modi{We also observed that in a given fraud case, cues rarely appear in isolation. Instead, certain cues frequently co-occur, for example, ``promise high returns with low risk,'' ``claim to be financial experts,'' and ``pitch senior-living investment schemes'' often appear together in one fraud case. Therefore, to better mirror real-world patterns, our intervention trains older adults using cue sets\footnote{\modi{A cue from one set may still combine with cues from other sets in real-world cases. Our cue sets only represent combinations commonly observed in our dataset, not the only possible groupings.}} rather than individual cues. Moreover, our goal is not to have participants learn every cue, but to assess the learning outcomes associated with different perspectives (Experiencer, Helper, Observer). Thus, to avoid overburdening older adults, we only selected six cues (i.e., two sets) per context. We chose two sets in each context so that subsequent evaluation, particularly the fraud cue identification test, would yield more robust results (see Sec. \ref{Online Fraud Awareness Assessment} for details). The selected fraud cues are shown in \autoref{online fraud cues}, with full fraud cues and their specific definitions in the Supplementary Material. All these cues were validated by one police officer who is working on the ground for anti-fraud education and three S\&P researchers, with more details in Sec. \ref{pilot study}.}

\begin{table}[h]
\caption{\modi{Online fraud cues.}}
\Description{Table 1: The online fraud cues within finance and health contexts.}
\centering
\begin{tabular}{@{} p{0.9cm}p{0.3cm}p{6.7cm} @{}}
\toprule
Context & Set & Cue \\
\midrule
\multirow{6}{*}{Finance}
  & \multirow{3}{*}{A}
    & Cue 1: Promise high returns with low risk \\
  &   & Cue 2: Claim to be financial experts \\
  &   & Cue 3: Pitch senior-living investment schemes \\
  \cmidrule{2-3}
  & \multirow{3}{*}{B}
    & Cue 4: Leverage peer influence \\
  &   & Cue 5: Invoke fake government investment schemes \\
  &   & Cue 6: Promise dividends between shareholders \\
\midrule
\multirow{6}{*}{Health}
  & \multirow{3}{*}{A}
    & Cue 1: Exaggerate illness to create anxiety \\
  &   & Cue 2: Claim to be medical experts \\
  &   & Cue 3: Advertise referrals to top doctors \\
  \cmidrule{2-3}
  & \multirow{3}{*}{B}
    & Cue 4: Frame products as miracle cures with refund guarantees \\
  &   & Cue 5: Offer free services as bait \\
  &   & Cue 6: Using cheap trials with steep follow-up charges \\
\bottomrule
\end{tabular}
\label{online fraud cues}
\end{table}


In addition, our analysis identified a variety of online fraud channels, such as instant messaging applications (e.g., WeChat and QQ) and livestreaming platforms, which are commonly used in online fraud. Accordingly, for the design of \modi{ROLESafe}, we selected a WeChat-style chat interface as the interaction medium to ensure both ecological validity and alignment with older adults' communication habits in China.

\subsection{Learning Mode Implementation}
\label{Learning Mode Implementation}
We chose LLM over other methods (e.g., Wizard-of-Oz) to implement these learning modes for both practical and conceptual reasons. Our long-term goal is to develop scalable tools for online fraud education, and using an LLM allows us to test an intervention format closer to real deployment. The study also involved multi-turn, role-based interactions across several conditions, making a human-wizard approach labor-intensive and prone to inconsistency and experimenter bias. This study serves as an initial demonstration of the feasibility of an LLM-based intervention, with future work needed to further refine dialogue design.
\subsubsection{Experiencer} 
We employed GPT-4o \cite{url9} as the backend model, which acted as a simulated scammer. First, to ensure the LLM understood our goal, we explicitly stated that its objective was to function as an online fraud simulation education system designed to enhance older adults’ ability to recognize fraud. We then specified the scammer role, instructing the LLM to play the role of a scammer attempting to defraud an elderly person on WeChat. To avoid any misunderstanding, we emphasized that this was strictly a simulation for educational purposes and not a real scam.

Moreover, we provided the set of online fraud cues and their definitions in \autoref{online fraud cues} to guide the scope of the LLM's responses. We established strict rules in this implementation. The model must use, and only use, the predefined online fraud cues and prohibit the introduction of any additional fraud cues to ensure that all cues are systematically covered and remain within our defined boundaries. To maintain the balance and naturalness of these cues, we instructed the model to balance the frequency and depth of each cue, while allowing the order of their appearance to flexibly adapt to the flow of the conversation.

Third, to create a realistic and user-friendly interaction, we specified the language style. The tone should be natural and realistic, responses were to be concise to avoid distracting older adults, and the total number of conversation turns was controlled to prevent feelings of fatigue or over-engagement. Additionally, we implemented a user-controlled exit option, allowing participants to end the conversation. Finally, we provided several example dialogues for the LLM to reference as guidance. The prompt used in this mode, coupled with few-shot learning \cite{brown2020language}, is provided in \modi{the Supplementary Material}. 

\subsubsection{Helper} GPT-4o served as our backend model to simulate a potential fraud victim. We also stated our educational goal. Then the LLM was instructed to simulate a potential victim by retelling to the user what the scammer said to it in a turn-taking dialogue, revealing fraud cues for older adults to identify. Moreover, the scammer's words that the LLM retold were strictly based on the predefined online fraud cues and their definitions in \autoref{online fraud cues}, and the same rules for cue use applied as in the Experiencer mode.

Since the objective of the Helper mode was to enable older adults to learn by assisting the victim in identifying fraud cues, we designed a mechanism that required the LLM to actively seek help from the user. For example, the LLM was instructed to phrase its responses in a help-seeking manner to elicit user reasoning. To ensure genuine recognition, we specified that vague advice such as ``don't trust it'' would not be considered successful identification. Given the LLM's strong capabilities, it was prohibited from directly teaching users how to recognize scams. To prevent discouragement, the LLM was further instructed to adapt its reactions when users explained the scam logic by displaying hesitation or partial doubt to encourage users to explain additional cues. 

Languages followed the natural and realistic style of the Experiencer mode. We also added a user-controlled exit option and provided example dialogue for guidance. The prompt is included in \modi{the Supplementary Material}. 

\subsubsection{Observer}
In this mode, older adults acted as third-party observers of simulated scam conversations. These conversations were generated using the extracted fraud cues described in \autoref{online fraud cues}. Specifically, we provided the fraud cues as part of the input to GPT-4o and requested it to generate scam dialogues, which were subsequently refined by S\&P experts to ensure plausibility. The prompt used for generating these dialogues is provided in \modi{the Supplementary Material}.

\begin{figure*}[htbp]
	\centering
		\includegraphics[scale=0.4]{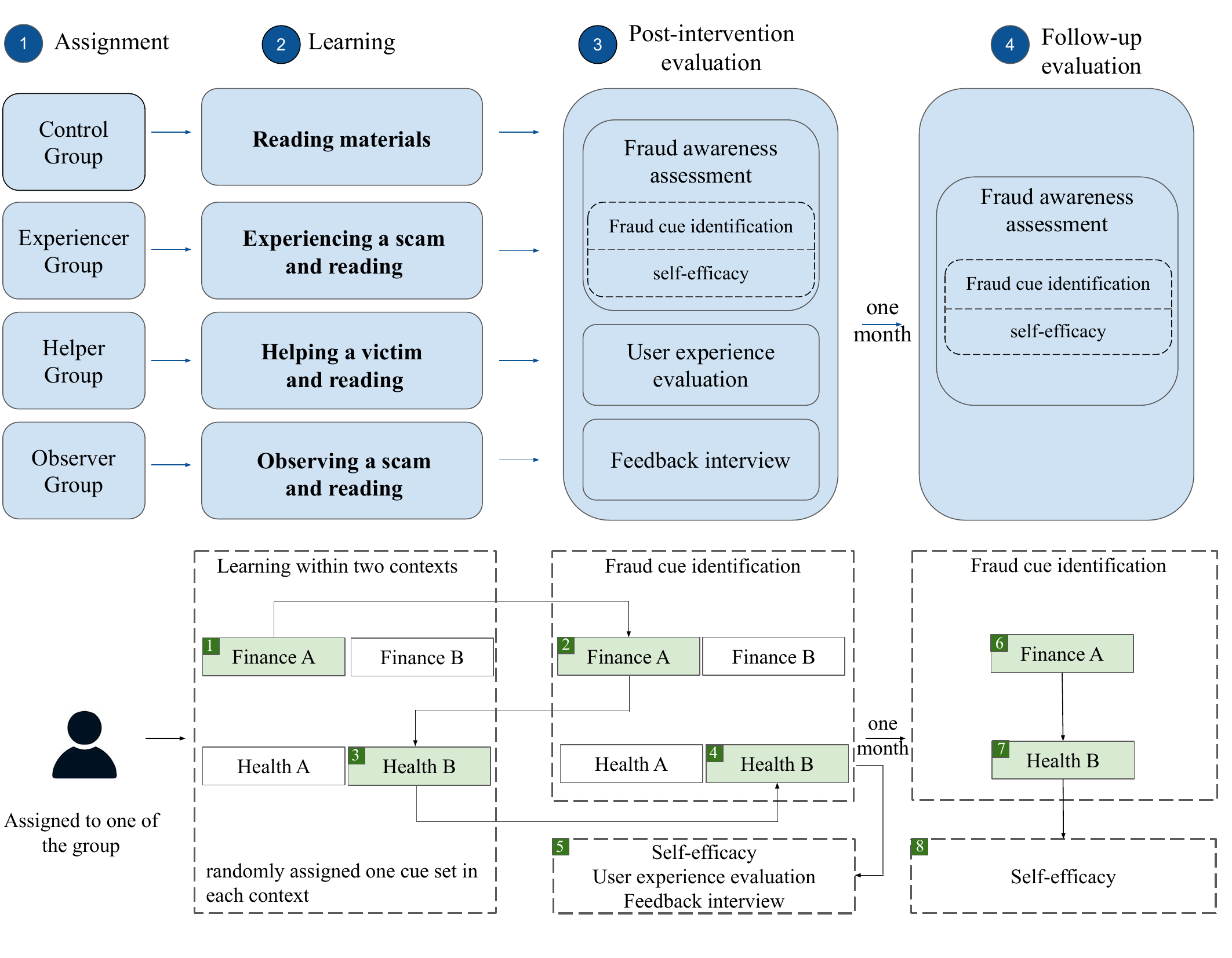}
	\caption{\modi{Study overview. The blue sections illustrate the overall study design. The green sections depict the study procedure for an individual participant. Note: participants were randomly assigned to begin with either the Finance or Health context, and the A/B cue sets were also randomly assigned.}}
    \Description{Figure 2: Study overview. For the study design, participants were recruited and assigned to one of four groups: control, experiencer, helper, or observer. Each group engaged in different learning activities, such as reading materials, experiencing or observing scams, or helping a victim. Post-intervention evaluation included fraud awareness assessment, user experience evaluation, and feedback interviews, followed by a one-month fraud awareness follow-up assessment. For the study procedure, each participant was first assigned to a group. They then learned the fraud cues in one context (e.g., finance A) and completed a fraud-cue identification test in the same context. Next, they learned the fraud cues in the other context (e.g., health B) and completed the corresponding identification test, followed by a self-efficacy scale, user-experience evaluation, and feedback interview. About one month later, the participant completed fraud-cue identification tests again in both contexts (e.g., finance A and health B) and repeated the self-efficacy scale.}
	\label{study_design}
\end{figure*}

\section{Study Procedure}
We conducted a between-subjects experiment to examine the effect of \modi{ROLESafe} on improving older adults' online fraud awareness.  \autoref{study_design} shows an overview of the study design.

\subsection{Pilot Study}
\label{pilot study}
Before the official data collection, we conducted two rounds of pilot studies with a total of 13 participants, \modi{including four experts (i.e., one police officer and three researchers) and nine older adults.} Participants were recruited through personal networks. Pilot sessions were conducted either via Zoom or in person, depending on participants' preferences. 

\modi{The experts primarily evaluated the study design and assessed the rigor and appropriateness of the materials, including the online fraud cues, the learning materials, and the GPT-generated fraud cases. The police officer brought extensive experience delivering anti-fraud education to diverse groups such as older adults and students. All researchers had backgrounds in security and privacy, with one having prior experience with online fraud research and another in phishing-related educational interventions. We refined our intervention design based on the collected feedback. For example, they helped refine the phrasing used in the fraud cases to better reflect real-world practices and noted the potentially high workload for older adults.}

The older adults focused on experiencing one of the four learning modes to provide feedback on study duration, workload, and overall feasibility. For example, they noted that the font size was too small on mobile devices, prompting us to enlarge the text and make the platform accessible across phones, tablets, and computers. 
\modi{We also observed that sessions lasting more than 30 minutes caused noticeable fatigue among older adults. Through iterative refinement and testing, we 
limited each context to six cues to be learned. With these adjustments, participants were able to complete all components in roughly 30 minutes on average while remaining attentive and engaged. Feedback indicated that participants did not feel rushed or perceive the tasks as overly demanding.} Data collected from the pilot studies were employed to improve our study design, and they were excluded from the final analysis.

\subsection{Participants and Recruitment}
We recruited Chinese participants who met the following inclusion criteria: 1) aged 50 or above, which aligns with the legal earliest retirement age in China \cite{retiree} and is consistent with prior research involving older adults \cite{he2023have,wan2019appmod,herbert2022fast}; 2) having basic communication abilities; and 3) \modi{using digital devices in their daily life (e.g., mobile phones and tablets).}\footnote{\modi{During the experiment, only technical assistance related to platform operation (e.g., navigating to the ``send'' button) was provided, as the interface was new to participants. No cognitive or decision-making assistance was given.}} A power analysis (power = 80\%, $\alpha$ = .05) for a one-way ANOVA conducted with G*Power suggested a sample size of 120 participants. To achieve this, we collaborated with a large local community center that regularly organizes activities for hundreds of older adults, which helped ensure the diversity of our sample. The community staff assisted in recruiting eligible participants.  

In the pilot study, we found that requiring older adults to complete a screening survey imposed additional cognitive demands. To avoid participant fatigue, we instead held a short informal conversation with each participant before the study began. This conversation served both to build rapport and to confirm inclusion criteria (e.g., age and experience with digital devices). As the community center pre-screened participants according to our criteria, only two individuals were later found ineligible for not meeting the age requirement. In total, we successfully recruited 150 participants. After excluding six participants with missing data, our final sample consisted of 144 older adults.

Regarding compensation, the collaborating community center advised that tangible items (e.g., eggs, rice, and bottled water) would be more culturally appropriate than direct monetary payments. Providing cash might affect older adults' expectations and reduce enthusiasm for future community events, since the center itself does not distribute money. Therefore, each participant received a gift valued at 50 RMB (approximately 6.96 USD) as compensation for their time and effort.\footnote{The compensation was determined with the aim of exceeding the hourly minimum wage of the province where participants are located (i.e., 21.0 yuan in Chongqing~\cite{url10}).}

Among the 144 participants, 52.08\% were female ($n$=75) and 47.92\% were male ($n$=69). Age distribution was as follows: 50–59 ($n$=78, 54.17\%), 60–69 ($n$=40, 27.78\%), 70–79 ($n$=21, 14.58\%), and 80+ ($n$=5, 3.47\%). Educational background included primary school or below ($n$=58, 40.28\%), middle school ($n$=38, 26.39\%), high school ($n$=10, 6.94\%), associate degree ($n$=25, 17.36\%), and bachelor’s degree or above ($n$=13, 9.03\%). We include the detailed demographic overview of the participants in \modi{the Supplementary Material}.

\subsection{Learning through ROLESafe}

While \modi{ROLESafe} was designed as a digital intervention, all sessions were in-person. \modi{We scheduled sessions with older adults in advance and conducted the intervention at the community center to ensure the setting felt familiar and comfortable for them.} Before the learning session began, we welcomed the older adults, introduced ourselves, explained the purpose of the project, collected basic demographic information, and obtained informed consent. As some participants might not have been familiar with the terminology of online fraud cues, we asked all participants to first read the same learning materials in \modi{the Supplementary Material} to establish a basic understanding before engaging in their assigned condition. The learning materials included descriptions of fraud cues, corresponding online fraud examples, and anti-fraud strategies, which were derived from the design materials in Sec. \ref{Design Materials} and validated by S\&P experts. 
\modi{Participants were then randomly assigned to one of the four conditions and instructed to complete tasks according to their assigned group. Participants chose the devices they prefered to complete the intervention, we provided iPad Air, Macbook Air, and iPhone 13. The interface displayed the same content across all devices and only the text size was adjusted according to participants' comfort.}


\begin{figure*}[htbp]
	\centering
		\includegraphics[scale=0.7]{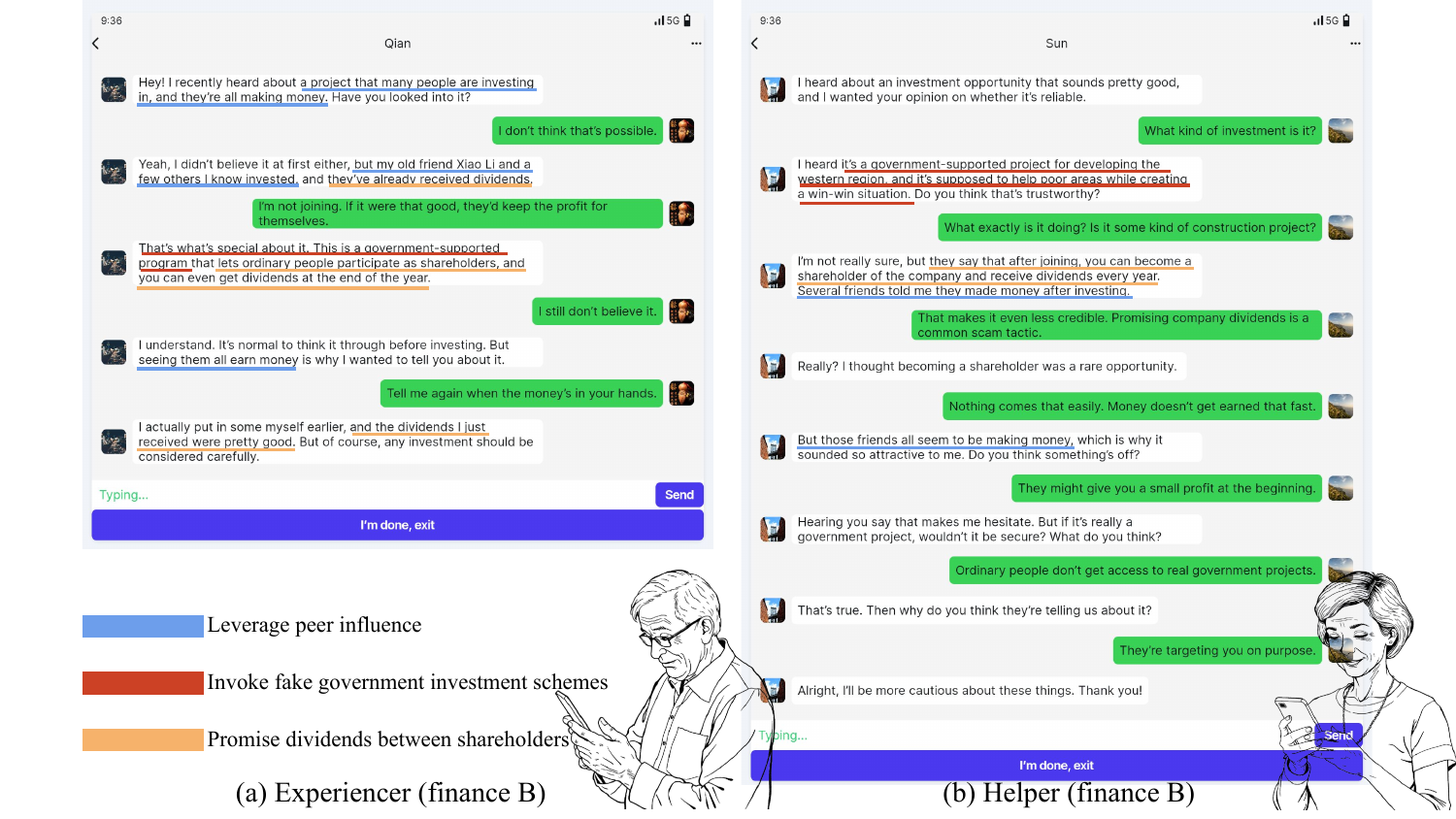}
	\caption{\modi{Interaction examples in the learning phase between users and LLM (taking the finance B context as an example). (a) An authentic conversation between an experiencer-group user and the LLM. (b) An authentic conversation between a helper-group user and the LLM. Colored highlights indicate how different fraud cues are manifested by the model.}}
    \Description{Figure 3: Examples of how users interacted with the LLM during the learning phase in the finance-B context. In the experiencer condition (left), the model plays the role of a scammer attempting to persuade the user, revealing fraud cues such as leveraging peer influence, invoking fabricated government-backed investment schemes, and promising shareholder dividends. In contrast, in the helper condition (right), the model adopts a supportive stance, guiding the user to question suspicious claims and recognize common fraud tactics.}
	\label{experiencer_helper_learning}
\end{figure*}

\begin{enumerate}
    \item Experiencer group: Participants were told that their task was to chat with another person through a WeChat-style interface. The conversation continued until they no longer wished to proceed. \modi{\autoref{experiencer_helper_learning}a shows a learning example between an experiencer-group user and the system.}
    \item Helper group: Participants were told that their task was to assist a potential victim of fraud by persuading the individual not to be deceived. They interacted with the victim through the WeChat-style interface, with the conversation lasting as long as they chose. \modi{\autoref{experiencer_helper_learning}b shows a learning example between a helper-group user and the system.}
    \item 
    Observer group: Their task was to observe chat records displayed in the WeChat-style interface. These generated chat records are shown in \modi{the Supplementary Material}.
    \item Control group: These older adults did not perform additional tasks. 
\end{enumerate}

During the intervention, researchers remained present throughout the session to accompany participants but did not interfere with the tasks. \modi{The detailed model performance is shown in Sec. \ref{llm performance}.}


\subsection{Post-Intervention Evaluation}

We conducted a series of post-intervention evaluations. System effectiveness and user experience are two critical evaluation dimensions in cybersecurity and privacy research \cite{huang2025systemization}. Effectiveness was evaluated using an online fraud awareness assessment, including fraud cue identification tests and a self-efficacy scale, while user experience was measured with a questionnaire-based evaluation. Finally, we gathered participants' perceptions and suggestions with exit interviews.

\subsubsection{Online Fraud Awareness Assessment.}
\label{Online Fraud Awareness Assessment}

\begin{figure*}[htbp]
	\centering
		\includegraphics[scale=0.65]{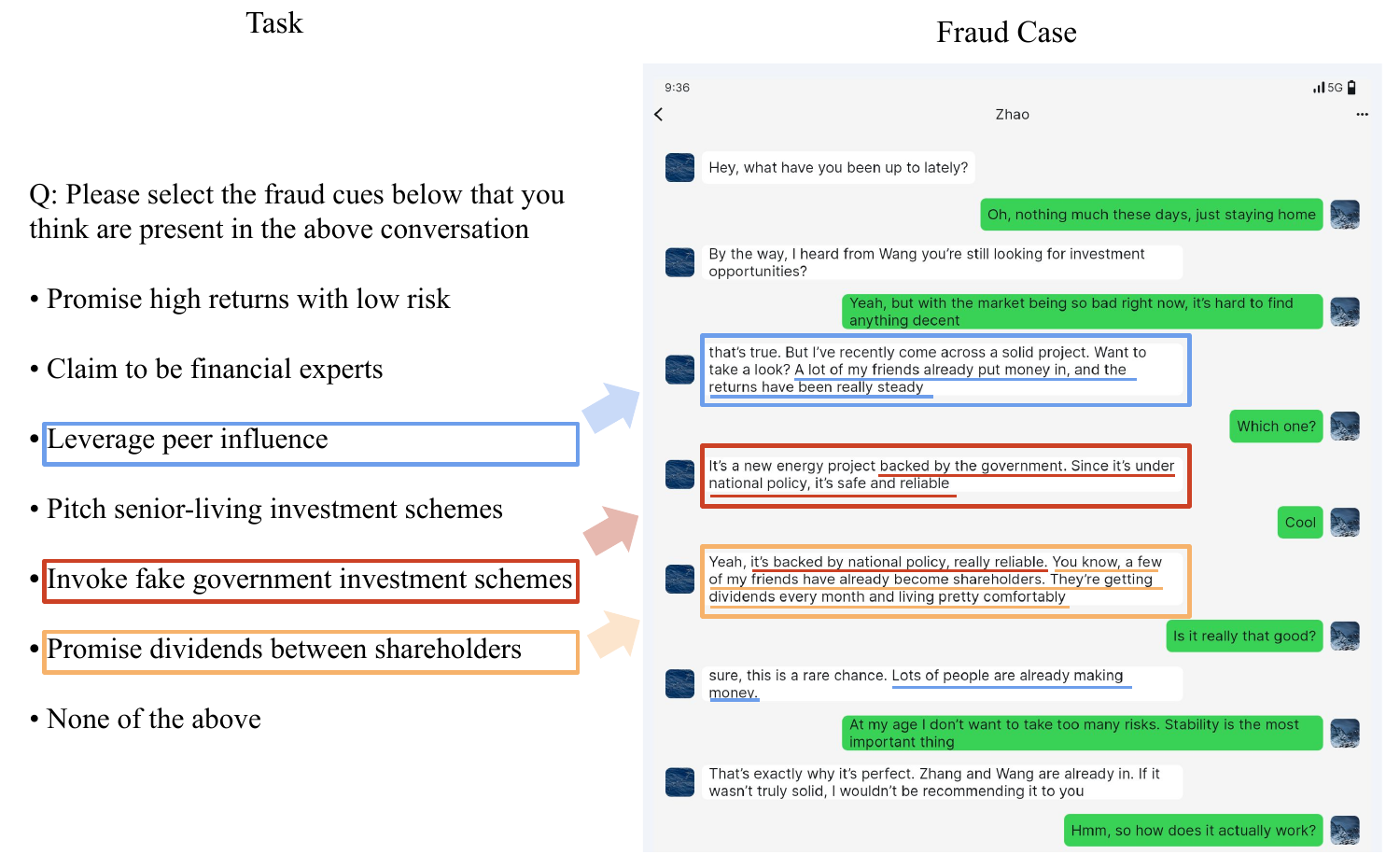}
	\caption{\modi{An illustration of a fraud cue identification test within a finance context. It demonstrates how specific fraud cues (left) correspond to suspicious elements in a fraud test case (right).}}
    \Description{Figure 4: Fraud-cue identification task used in the finance context. Participants were shown a conversational fraud case (right) and asked to select which cues were present from a predefined list (left). The highlighted segments in the conversation map onto specific fraud cues, such as leveraging peer influence, invoking fabricated government-backed projects, and promising shareholder dividends, demonstrating how these cues were embedded in the scenario and surfaced for recognition during the test.}
	\label{fraud cue identification example}
\end{figure*}

\textit{Fraud cue identification test.} \modi{ \autoref{fraud cue identification example} shows an example of a fraud cue identification test, including a fraud case and a task.\footnote{\modi{We did not include non-scam cases because our goal was to assess whether participants could identify targeted fraud cues, rather than to evaluate binary classification performance and adding non-scam cases would shift the construct with limited benefit for the training objective.}} Participants were presented with an online fraud test case 
before completing the cue identification task.  
The task was ``Please select the fraud cues below that you think are present in the above conversation.'' Participants selected any number of cues they believed appeared in the case from six possible fraud cues that had been identified for the given context.} 

\modi{\autoref{study_design} (green parts) illustrates how a participant goes through the learning and evaluation phases. Specifically, after completing the learning phase in one context, participants took a fraud cue identification test corresponding to that context (e.g., finance context). They then proceeded to the second learning phase in the other context and completed another test specific to that context (e.g., health context). Thus, each participant completed two fraud cue identification tests in total. The order of fraud contexts in the learning phases was randomized, and within each context, the tested cue set (A or B) was also randomly assigned. For instance, a participant might work with cue set A in the financial context and cue set B in the health context. This design ensured that participants did not all encounter the same cue set, thereby improving the robustness of the tests.}

\modi{We adopted a multiple-choice format rather than a per-cue binary response to reduce satisficing behavior observed in the pilot study, where some older participants tended to endorse all cues without careful consideration. The multiple-choice layout encourages comparative thinking and deliberate selection. To evaluate identification accuracy, we used the F1 score, which balances precision and recall, providing a relatively fair measure of overall cue-recognition ability \cite{powers2020evaluation}. We computed the F1 score by taking twice the number of options that were selected in both the predicted and ground-truth answers, and dividing it by the total number of options selected across the predicted and the ground-truth answers. It ranges from 0 to 1, with higher values indicating stronger fraud-cue recognition ability. The final F1 score was calculated as the average of the two tests.}

 


The test cases were generated using the fraud cues collected in Sec. \ref{fraud cue}. Specifically, each of the four fraud cue sets from \autoref{online fraud cues} was provided as part of the input prompts to GPT-4o, which generated four online fraud test cases. Researchers then refined the outputs to ensure plausibility. 
The prompt used for case generation is provided in \modi{the Supplementary Material}, and the four complete test cases are provided in the Appendix \ref{Cases in Fraud Awareness Assessment (Main Study)}.


\textit{\modi{Self-efficacy}.} In addition to the fraud cue identification tests, we also used a self-efficacy scale to assess older adults' confidence in identifying online fraud. After two fraud cue identification tests were completed, participants were guided to assess their self-efficacy, with measures adapted from \cite{ormond2016perceived}. The specific items were shown in the Appendix \ref{scale question}. The system logged the responses automatically. 

\subsubsection{User Experience Evaluation}
\label{User Experience}
After finishing the online fraud awareness assessment, participants were asked to report their learning experience with the assigned intervention. This evaluation was conducted through a survey (details in Appendix \ref{scale question}). The survey was implemented on our application following the awareness tests, and participants' responses were automatically logged.  

Specifically, we employed three constructs that are widely used to evaluate learners' learning experiences with interventions in cybersecurity education research, including enjoyment \cite{hull2023tell,ye2025awareness}, perceived usability \cite{alqahtani2020design,filipczuk2019using}, and future behavioral intention \cite{ebbers2016authentication,denning2013control}. 
The three constructs were measured as follows: 
\begin{enumerate}
    \item \textit{Enjoyment}: Participants' perceived pleasure and fulfillment when engaging with the assigned online fraud education, adapted from \cite{agarwal2000time}. 
    \item 
    \textit{Usability}: Participants' general perceptions of effectiveness, efficiency, and satisfaction of the learning system, adapted from the Usability Metric for User Experience (UMUX) \cite{finstad2010usability}.
    \item \textit{Future behavioral intention}: Participants' intention to engage with the assigned education again in the future, adapted from \cite{agarwal2000time}. 
\end{enumerate}
Two researchers reviewed and adapted these scales from English to Chinese. Each item was measured using a 5-point Likert scale, with response options ranging from ``strongly disagree'' to ``strongly agree.'' The detailed user experience questions are shown in the Appendix \ref{scale question}.
\subsubsection{Intervention Feedback Interview}
The interview covered several themes related to participants' experiences and perceptions. For example, we asked about: 1) prior experiences with anti-fraud education, 2) their perceptions of the assigned role in the intervention, 3) impressions of the interaction partner (e.g., whether they felt it was a human or a machine, particularly for participants in the Experiencer and Helper groups), 4) overall learning perceptions, 5) perceived task load, and 6) anticipated impact of the intervention on their future behavior. The full list of interview questions is provided in \modi{the Supplementary Material}. Each session, including the learning phase, tests, survey, and interview, lasted approximately 30 minutes in total.

\subsection{Follow-up Intervention Evaluation}

We reconnected with the collaborating community center, which assisted us in inviting participants who had taken part in the main study to voluntarily return for a follow-up session approximately one month later.\footnote{The timing was around one month rather than an exact day, as the main study itself spanned several weeks, and we could not ensure that each participant returned on precisely the same day.} 
A total of 89 participants completed the follow-up study, of which five were excluded due to repeated tests, resulting in 84 valid users, including 17 participants in the Experiencer group, 31 participants in the Helper group, 17 participants in the Observer group, and 19 participants in the Control group. The demographics of these older adults is shown in \modi{the Supplementary Material}.


In the follow-up session, participants completed the same two measures as in the main study: 1) the fraud cue identification test and 2) the self-efficacy scale, both serving as indicators of intervention effectiveness. While the tests targeted the same fraud cues as before, the test cases were newly generated following the same procedure described in Sec. \ref{Online Fraud Awareness Assessment}. The context order and test sets for each participant were consistent with their allocation in the main study. \modi{The four complete test cases for follow-up evaluation are provided in the Appendix \ref{Cases in Fraud Awareness Assessment (Follow-up Study)}.} Each follow-up session lasted approximately 10 minutes. Participants who took part in this additional session received an extra gift valued at 15 RMB (2.1 USD).

\subsection{Ethics Considerations}

This study received approval from our Institutional Review Board (IRB). Because our research introduced GPT-4o with customized prompts to interact with older adults, we implemented several safeguards to minimize the risk of unintended outputs beyond our control. First, in the prompt, we try our best to highlight the system's educational purpose and to restrict the content strictly to the predefined online fraud cues, as shown in Sec. \ref{Learning Mode Implementation}. Second, although these safeguards significantly reduced risk, we could not guarantee with absolute certainty that no unexpected content would be generated. Therefore, all interventions were conducted in person, with a researcher present throughout the session. If GPT-4o produced, or showed signs of producing, unintended content, the session would be immediately terminated. Across all participants in the main study and the pilot sessions, no such incidents occurred. 

Moreover, given the sensitivity of fraud, all participants were clearly informed that the intervention was conducted for educational purposes on online fraud. Although in the consent process, we withheld specific information about the use of an LLM-based simulation to prevent bias or priming prior to the intervention, participants were fully briefed about the role of LLMs and invited to share their perception in the feedback interview.

To further protect participants' privacy, we only collected basic demographic information (age, gender, and education). Recruitment was conducted through a community center to minimize direct contact with participants. If any personal information was mentioned in transcripts, we removed it during data processing.


\section{Data Analysis}

\subsection{Quantitative Data Analysis}

The quantitative data included responses to the online fraud awareness assessment in Sec. \ref{Online Fraud Awareness Assessment} and the user experience questionnaire in Sec. \ref{User Experience}.  

\subsubsection{Randomization Check.} We first examined data completeness and excluded cases with missing values. Randomization of group assignment was then assessed across demographic factors, including age, gender, and education level. For age, normality (Shapiro–Wilk test) and homogeneity of variance (Levene's test) were evaluated. As the assumptions for ANOVA were not met, the non-parametric Kruskal–Wallis test was applied. Gender distribution was examined using a chi-square test of independence. Education level was assessed using Fisher's exact test with Monte Carlo simulation due to sparse cells (expected counts smaller than five) as an alternative to the chi-square test for larger contingency tables.

\subsubsection{Data Processing.} For scale-based data, including the self-efficacy scale and user experience questionnaires, negatively worded items were reverse scored.  To examine the validity and reliability of our measurement scales, we employed both internal consistency estimates and confirmatory factor analysis (CFA). Specifically, Cronbach’s $\alpha$ and McDonald’s $\omega$ were computed for each multi-item construct to assess internal consistency. CFA was then conducted to examine the factor structure and construct validity, with standardized factor loadings and model fit indices (e.g., $\chi^{2}$/df, CFI, TLI, RMSEA, SRMR) reported. After confirming acceptable reliability and validity, we calculated the mean score of each construct for each participant, which served as the indicator of their overall level on that construct. 


\subsubsection{Difference between Groups.}
\label{Difference between groups}
To evaluate the overall impact of the role-based simulation intervention on effectiveness and user experience, we compared the combined treatment groups (Experiencer, Helper, Observer) against the control group using the Mann–Whitney U test \cite{conover1999practical}. This non-parametric test was chosen because the Shapiro–Wilk test indicated that the data were not normally distributed, and the two groups are independent.  

To further examine group-level differences, we conducted two sets of Kruskal–Wallis H tests \cite{kruskal1952use}. First, we compared outcomes across all four groups (Experiencer, Helper, Observer, Control) to identify differences between the treatment and control groups. Second, we compared outcomes across the three treatment groups (Experiencer, Helper, Observer) to investigate any role-specific effects. This non-parametric test is appropriate for independent groups when the dependent variable is ordinal or continuous but not normally distributed. When the Kruskal–Wallis test indicated significance, we conducted post-hoc pairwise comparisons using Dunn's test with Bonferroni correction to adjust for multiple testing. 


\subsubsection{Follow-up Study Analysis.} 
To examine whether the effectiveness of the intervention was retained over time in each group, we compared participants' fraud cue identification and self-efficacy between the main study and the follow-up. When data were approximately normally distributed, we used paired $t$-tests; otherwise, we applied Wilcoxon signed-rank tests.  
We applied the same methods to compare group differences in the follow-up evaluation as in the main study, shown in Sec. \ref{Difference between groups}.

\subsection{Qualitative Data Analysis}

The qualitative data included feedback interviews. All qualitative data were collected as audio recordings. The recordings were first transcribed into text for subsequent analysis.  
We conducted an inductive thematic analysis \cite{braun2006using}. Specifically, two authors independently conducted open coding \cite{corbin2014basics} on a random sample of 40 transcripts to familiarize themselves with the data and to generate preliminary codes. This stage emphasized capturing insights directly emerging from the data. Examples of codes included ``having a sense of achievement'' and ``having a deeper impression.'' The researchers then iteratively discussed and compared their codes through multiple rounds of meetings to resolve disagreements and ensure consistency, constructing an initial codebook. Then, two authors independently coded half of the remaining transcripts, where we revisited and refined the codebook through multiple meetings. Subsequently, we applied affinity diagramming \cite{muller2014curiosity} to cluster codes with similar meanings into higher-level themes by examining the codes, their associated excerpts, and the relationships among them. Examples of higher-level themes included ``the positive perceptions of LLM-based simulation'' and ``the negative perceptions of Helper.'' \modi{Due to our participants are Chinese older adults, all the transcripts are in Chinese. All quotes were first translated from Chinese to English using Google Translate. The first author then reviewed and refined the translations to ensure their accuracy.}

\section{Findings}
\subsection{Preliminary Validation Analysis}
There were 33 participants in the Experiencer group, 40 participants in the Helper group, 35 participants in the Observer group, and 36 participants in the Control group. Randomization check confirmed that older adults were evenly distributed across groups. Specifically, the chi-square test indicated no significant differences in gender proportions ($\chi^{2}$(3)=0.50, \textit{P}=.920, \text{Cramér's V}=0.059), Fisher's exact test with Monte Carlo simulation showed no significant differences in education level ($\textit{P}$$ \approx$$.276$, \text{Cramér's V}=0.181), and the Kruskal–Wallis test indicated that age differences among the four groups did not reach statistical significance (H(3)=7.56, \textit{P}=.056).

The results of a CFA indicated an acceptable model fit ($\chi^{2}$/df=2.06, CFI=0.915, TLI=0.894, RMSEA=0.086, SRMR=0.072). All constructs exhibited Cronbach's alpha values near or above 0.7, demonstrating acceptable reliability. Two constructs, usability and self-efficacy, had slightly lower values of 0.634 and 0.666, respectively. While 0.7 is often cited as the benchmark for acceptable internal consistency, recent methodological reviews note that values approaching this threshold can still be considered adequate, particularly in exploratory or applied contexts \cite{taber2018use}. Given that both values are close to 0.7, we retained these constructs in the analysis.

\subsection{Effectiveness of Intervention}
\subsubsection{Online Fraud Cue Identification Comparison.}
\label{Online Fraud Cue Identification Comparison}
\begin{figure*}[htbp]
    \centering
    \begin{subfigure}{0.5\textwidth}
        \centering
        \includegraphics[width=\linewidth]{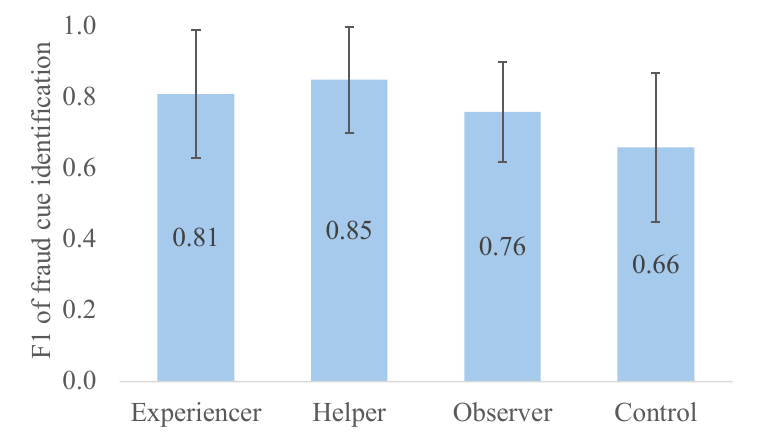}
        \caption{Fraud cue identification (mean ± SD)}
        \label{f1_main}
    \end{subfigure}
\hspace{-0.04\textwidth}
    \begin{subfigure}{0.5\textwidth}
        \centering
        \includegraphics[width=\linewidth]{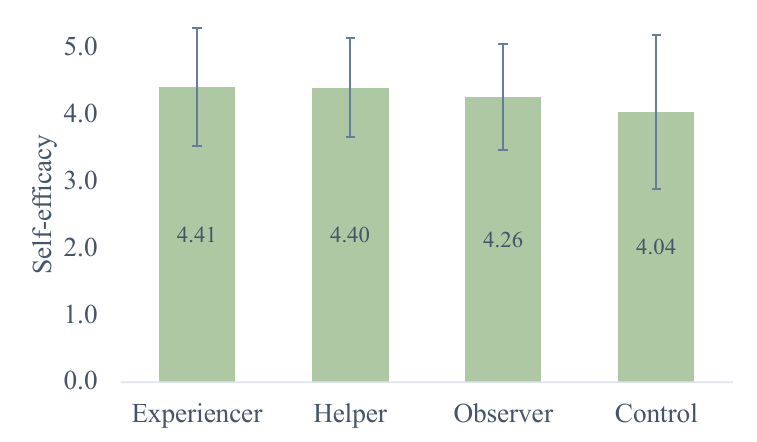}
        \caption{Self-efficacy (mean ± SD)}
        \label{SE_main}
    \end{subfigure}
    
    \caption{Descriptive comparisons of effectiveness across groups.}
    \Description{Figure 5: Bar charts showing descriptive comparisons across groups. (a) Fraud cue identification (F1 scores) was higher in the treatment groups (Experiencer 0.81, Helper 0.85, Observer 0.76) than in the Control group (0.66). (b) Self-efficacy scores were similar across groups, with Experiencer (4.41), Helper (4.40), Observer (4.26), and Control (4.04).}
\end{figure*}

As shown in \autoref{f1_main}, participants in all treatment groups (Experiencer, Helper, Observer) achieved higher average $F_1$ scores in the online fraud cue recognition compared to the control group. A Mann–Whitney U test confirmed that these differences were statistically significant (W=1009.5, \textit{P}\textless0.001). A Kruskal–Wallis test revealed significant differences across the four groups (H(3)=24.68, \textit{P}\textless0.001), shown in \autoref{kw_main}. Post-hoc pairwise comparisons using Dunn's test with Bonferroni correction identified a significant difference between the Helper and Control group (Z=4.73, adjusted \textit{P}\textless.001), as well as between the Experiencer and Control group (Z=3.53, adjusted \textit{P}=.002), displayed in \autoref{dunn_all}.

\begin{table}[htbp]
\centering
\caption{Results of the Kruskal–Wallis test comparing intervention effectiveness.}
\Description{Table 2: Results of Kruskal–Wallis tests on intervention effectiveness. Compared to the control group, treatment groups showed significantly higher fraud cue identification (p $<$ .001), but no significant difference in self-efficacy. Among treatment groups, fraud cue identification also differed significantly (p = .031), while self-efficacy showed no significant difference.}
\begin{tabularx}{\columnwidth}{@{} p{1.5cm} p{3cm} c c @{}}
\toprule
 & Effectiveness Construct & Test Statistic & P-value \\
\midrule
\multirow{2}{*}{\makecell[l]{treatment\\ vs.\ control}}
 & fraud cue identification & 24.68 & $<.001^{***}$ \\
 & self-efficacy & 1.98 & .576 \\
\midrule
\multirow{2}{*}{\makecell[l]{among\\ treatment}}
 & fraud cue identification & 6.97 & $.031^{*}$ \\
 & self-efficacy & 1.09 & .581 \\
\bottomrule
\end{tabularx}
\label{kw_main}
\end{table}

\begin{table}[htbp]
\centering
\caption{Pairwise comparisons across all groups on fraud cue identification using Dunn's test with Bonferroni correction}
\Description{Table 3: Pairwise comparisons of fraud cue identification across groups using Dunn’s test with Bonferroni correction. Both the Helper group (p $<$ .001) and Experiencer group (p = .002) performed significantly better than the Control group.}
\begin{tabular}{lll}
\toprule
               & Test Statistic & Adjusted P\\
\midrule
control vs. helper       & 4.73           & \textless{}.001***     \\
control vs. experiencer  & 3.53           & .002**     \\
control vs. observer     & 2.21           & .163         \\
experiencer vs. helper   & 1.00           & 1.000         \\
observer vs. experiencer & 1.35           & 1.000          \\
observer vs. helper      & 2.43           & .090         \\
\bottomrule
\end{tabular}

\label{dunn_all}
\end{table}

Among the three treatment conditions, the Helper group achieved the highest mean $F_1$ score, followed by the Experiencer group, while the Observer group had the lowest in \autoref{f1_main}. The Kruskal–Wallis test indicated a significant difference among the three treatment groups (H(2)=6.97, \textit{P}=.031) in \autoref{kw_main}. Dunn's post-hoc tests with Bonferroni adjustment showed that Helpers outperformed Observers (Z=2.63, adjusted \textit{P}=.025), whereas no significant differences were observed between Helpers and Experiencers or between Experiencers and Observers, shown in \autoref{dunn_treatment}.

\textbf{Summary:} 
Our findings show that role-based simulation learning proved more effective than the control condition in improving fraud cue recognition. Both Experiencers and Helpers outperformed the Control group, while Helpers further outperformed Observers. In other words, engaging older adults in more active roles, particularly helping others, may provide stronger benefits.

\begin{table}[htbp]
\centering
\caption{Pairwise comparisons among treatment groups on fraud cue identification using Dunn’s test with Bonferroni correction.}
\Description{Table 4: Pairwise comparisons among treatment groups on fraud cue identification using Dunn’s test with Bonferroni correction. The Helper group significantly outperformed the Observer group (p = .025). No significant differences were found between Experiencer and Helper or between Observer and Experiencer.}
\begin{tabular}{lll}
\toprule
Comparison               & Test Statistic & Adjusted P \\
\midrule
observer vs. helper      & 2.63           & .025*     \\
observer vs. experiencer & 1.50           & .399 \\
experiencer vs. helper   & 1.04           & .891  \\
\bottomrule
\end{tabular}
\label{dunn_treatment}
\end{table}

\subsubsection{Self-efficacy of Online Fraud Identification.}

    
All three treatment groups showed higher mean self-efficacy scores than the control group, as shown in \autoref{SE_main}. Nonetheless, the differences were not statistically significant, as indicated by both the Mann–Whitney U test (W=1737.5, \textit{P}=.314) and the Kruskal–Wallis test (H(3)=1.98, \textit{P}=.576) in \autoref{kw_main}.  
The mean self-efficacy scores were 4.41 for Experiencer, 4.40 for Helper, and 4.26 for Observer in \autoref{SE_main}. However, the Kruskal–Wallis test revealed no significant differences across these groups (H(2)=1.09, \textit{P}=.581) in \autoref{kw_main}.  

\textbf{Summary}: Taken together, these findings suggest that while role-based simulation learning may enhance older adults' self-efficacy in recognizing fraud, neither the presence of the role-based intervention nor the role assigned led to statistically significant improvements in self-efficacy.

\subsection{User Experience of Intervention}

\begin{figure*}[htbp]
    \centering
    \begin{subfigure}{0.33\textwidth}
        \centering
        \includegraphics[width=\linewidth]{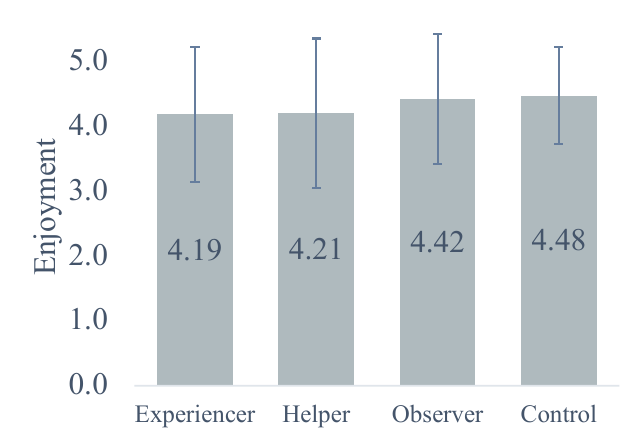}
        \caption{Enjoyment (mean ± SD)}
        \label{enjoyment}
    \end{subfigure}
\hspace{-0.02\textwidth}
    \begin{subfigure}{0.33\textwidth}
        \centering
        \includegraphics[width=\linewidth]{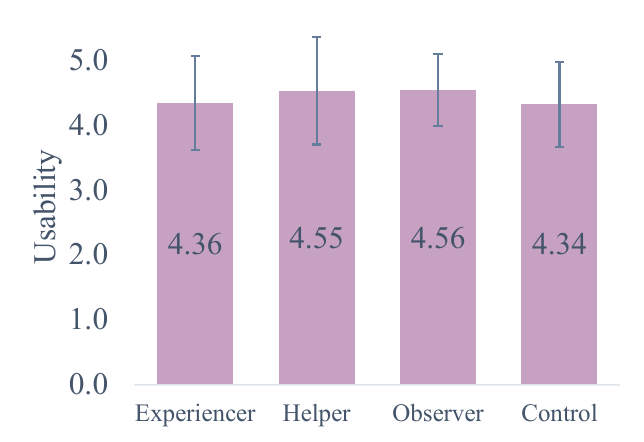}
        \caption{Usability (mean ± SD)}
        \label{usability}
    \end{subfigure}
    \hspace{-0.02\textwidth}
    \begin{subfigure}{0.33\textwidth}
        \centering
        \includegraphics[width=\linewidth]{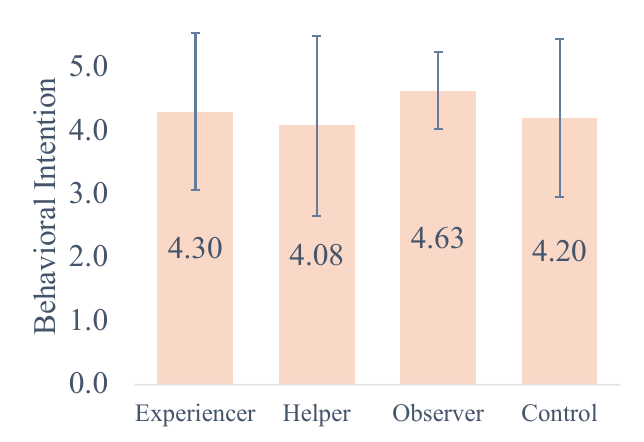}
        \caption{Future Behavioral Intention (mean ± SD)}
        \label{BI}
    \end{subfigure}
    
    \caption{Descriptive comparisons of user experience across groups.}
    \Description{Figure 6: Bar charts comparing user experience across groups. (a) Enjoyment scores were similar across groups, ranging from 4.19 (Experiencer) to 4.48 (Control). (b) Usability ratings were also comparable, with means from 4.34 (Control) to 4.56 (Observer). (c) Future behavioral intention scores ranged from 4.08 (Helper) to 4.63 (Observer), with no large differences between groups.}
    \label{user_experience}
\end{figure*}

\autoref{user_experience} presents the descriptive results for user experience measures. Overall, the Mann–Whitney U tests revealed no statistically significant differences between the treatment groups and the control group across any of the constructs. Likewise, the Kruskal–Wallis tests indicated no significant differences among the four groups (see \autoref{kw_user_experience}). Although the Observer group showed the highest means for \textit{Usability} and \textit{Future Behavioral Intention}, neither of these differences reached statistical significance. 

\begin{table}[htbp]
\centering
\caption{Results of the Kruskal–Wallis test comparing user experience.}
\Description{Table 5: Results of Kruskal–Wallis tests on user experience constructs. No significant differences were found between treatment and control groups or among treatment groups for enjoyment, usability, or future behavioral intention.}
\begin{tabularx}{\columnwidth}{@{} p{1.3cm} p{3.5cm} c c @{}}
\toprule
 & User Experience Construct   & Test Statistic & P-value \\
\midrule
\multirow{3}{*}{\makecell[l]{treatment\\ vs.\ control}} 
 & enjoyment                   & 2.21           & .529    \\
 & usability                   & 5.30           & .151   \\
 & future behavioral intention & 1.49           & .685   \\
\midrule
\multirow{3}{*}{\makecell[l]{among\\ treatment}}       
 & enjoyment                   & 1.80           & .406   \\
 & usability                   & 2.20           & .332   \\
 & future behavioral intention & 0.78           & .675   \\
\bottomrule
\end{tabularx}
\label{kw_user_experience}
\end{table}


\textbf{Summary:} Role-based simulation interventions did not yield significant differences in user experience, and the assigned roles also had no measurable impact on participants' perceptions.

\subsection{User-System Interaction}
\label{llm performance}
\modi{In this section, we describe the interaction statistics between users and ROLESafe. \autoref{cue coverage} demonstrates the cue coverage rates across groups and contexts during the learning phase. For the experiencer and helper groups, since they interact with the LLM directly, the entire process remains dynamic, so we manually annotated the cue coverage. Specifically, if the LLM produced an utterance that included a particular fraud cue, we considered that conversation to have covered that cue. Conversely, if none of the LLM's responses contained the cue, the conversation was coded as not covering it. For example, in \autoref{experiencer_helper_learning}a, the LLM's statement ``this is a government-supported program'' is counted as covering the fraud cue of ``invoke fake government investment schemes'' within that conversation. To establish a shared coding standard, two authors annotated the presence or absence of covered cues. To assess inter-rater reliability, the two authors independently coded a random sample of 70 conversations. The results showed
a Cohen's Kappa of at least 0.9 for every cue, indicating strong agreement between coders. Afterward, the coders discussed discrepancies and reached consensus on the annotations. Each author then individually coded half of the remaining conversations.
For the observer group, participants read pre-generated fraud cases rather than interacting with the LLM, so each fraud case includes all three cues.
}
 
\modi{For the experiencer and helper groups, LLMs provided all three cues in 65.07\% of user–LLM interactions, with 86.30\% containing at least two cues and 98.63\% containing at least one. We observed that 34.93\% (n=51) interactions did not cover all three cues. Most of these cases fell into two categories: 1) users signaled clear disinterest, prompting the LLM to refrain from pushing the conversation further, and the interaction then ended naturally by the user (21.92\%; n=32; n(helper)=3, n(experiencer)=29). 2) users abruptly ended the interaction before the LLM had finished providing all fraud cues (11.64\%; n=17; n(helper)=4, n(experiencer)=13). Only two interactions (1.37\%, n(helper)=1, n(experiencer)=1) failed to include the full cue set due to incomplete cue presentation by the LLM itself. We also found that the helper group exhibited substantially higher cue coverage rates than the experiencer group. One plausible explanation was that older adults tended to be more engaged in the helper role, whereas in the experiencer role, particularly within health contexts, they were less willing to sustain interaction with the system. This may shed light on why helpers demonstrate stronger fraud cue identification learning gains compared to experiencers, as reported in Sec. \ref{Online Fraud Cue Identification Comparison}.
}

\begin{table}[htbp]
\caption{\modi{The proportion of conversations that contained specific numbers of cues. A ``conversation'' refers to one complete chat log during a learning phase. For each metric (Cue ($n=3$), Cue ($n\geq2$), Cue ($n\geq1$)), the denominator is the total number of conversations within the corresponding group or context. The numerator is the number of conversations that covered exactly three cues, at least two cues, or at least one cue, respectively. Thus, for example, ``Cue ($n=3$)'' represents the percentage of conversations in which all three target cues appeared. Due to the conversation content in the observer group being fixed, we only combine the data for the other two groups to calculate the overall cue coverage rates of user-LLM interactions.}}
\Description{Table 6: Cue coverage rates across groups and contexts.}
\label{cue coverage}
\begin{tabular}{p{2cm}p{2cm}p{1cm}p{1cm}p{1cm}}
\toprule
Group                               & Context       & Cue ($n=3$) & Cue ($n\geq2$) & Cue ($n\geq1$) \\
\midrule
\multirow{3}{*}{Experiencer}        & finance       & 48.48\%   & 93.94\%                & 100.00\%               \\
                                    & health        & 21.21\%   & 57.58\%                & 100.00\%               \\
                                    & all           & 34.85\%   & 75.76\%                & 100.00\%               \\
\midrule
\multirow{3}{*}{Helper}             & finance       & 90.00\%   & 95.00\%                & 95.00\%                \\
                                    & health        & 90.00\%   & 95.00\%                & 95.00\%                \\
                                    & all           & 90.00\%   & 95.00\%                & 97.50\%                \\
\midrule
\multirow{3}{*}{Observer}           & finance       & 100.00\%  & 100.00\%               & 100.00\%               \\
                                    & health        & 100.00\%  & 100.00\%               & 100.00\%               \\
                                    & all           & 100.00\%  & 100.00\%               & 100.00\%               \\
\midrule
\multicolumn{2}{l}{Combined (Experiencer + Helper)} & 65.07\%   & 86.30\%                & 98.63\% \\
\bottomrule
\end{tabular}
\end{table}

\modi{We also present the interaction metrics in \autoref{engagement statistics}. For the experiencer group and helper group, on average, conversations contained 9.55 turns\footnote{Each message from either the user or the LLM is counted as one turn.}, lasted 4.40 minutes, and included 27.64 Chinese characters per turn. Overall, user–LLM engagement was higher in the Helper group than in the Experiencer group.}

\begin{table}[htbp]
\caption{\modi{User-system interaction metrics (mean(std)). Turn Count refers to the number of turns within a conversation. Time (min) reflects the duration of each conversation. Character Count captures the number of characters per turn. We do not have the observers' reading-time data.}}
\Description{Table 7: User-system interactions metrics including turn count, time, and character count.}
\label{engagement statistics}
\begin{tabularx}{\columnwidth}{@{} p{1.3cm}p{1.4cm}p{1.5cm}p{1.5cm}p{1.6cm}@{}}
\toprule
Group                               & Context       & Turn Count   & Time (min)       & Character Count \\
\midrule
\multirow{3}{*}{Experiencer}        & finance       & 7.82 (3.45)  & 3.18 (2.39)      & 25.37 (18.37)   \\
                                    & health        & 7.79 (3.71)  & 2.97 (1.85)      & 25.61 (18.48)   \\
                                    & all           & 7.80 (3.56)  & 3.08 (2.12)      & 25.49 (18.41)   \\
\midrule
\multirow{3}{*}{Helper}             & finance       & 11.03 (4.04) & 5.72 (3.14)      & 29.65 (19.21)   \\
                                    & health        & 10.98 (2.90) & 5.27 (2.74)      & 28.10 (18.16)   \\
                                    & all           & 11.00 (3.49) & 5.49 (2.93)      & 28.88 (18.70)   \\
\midrule
\multirow{3}{*}{Observer}           & finance       & 15.50 (0.71) & \textbackslash{} & 19.74 (14.38)   \\
                                    & health        & 12.00 (0.00) & \textbackslash{} & 18.42 (12.25)   \\
                                    & all           & 13.75 (2.06) & \textbackslash{} & 19.16 (13.38)   \\
\midrule
\multicolumn{2}{l}{\makecell[l]{Combined \\(Experiencer+Helper)}} & 9.55 (3.86)  & 4.40 (2.86)      & 27.64 (18.66)  \\
\bottomrule
\end{tabularx}
\end{table}

\subsection{One-month Follow-up Effects of Intervention}
\label{Short-effects of Intervention}
    
\begin{figure*}[htbp]
    \centering
    \begin{subfigure}{0.5\textwidth}
        \centering
        \includegraphics[width=\linewidth]{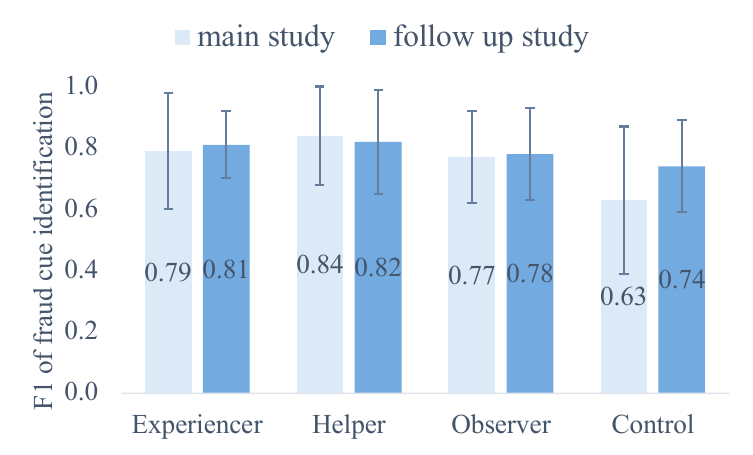}
        \caption{Fraud cue identification (mean ± SD)}
        \label{f1_follow}
    \end{subfigure}
\hspace{-0.04\textwidth}
    \begin{subfigure}{0.5\textwidth}
        \centering
        \includegraphics[width=\linewidth]{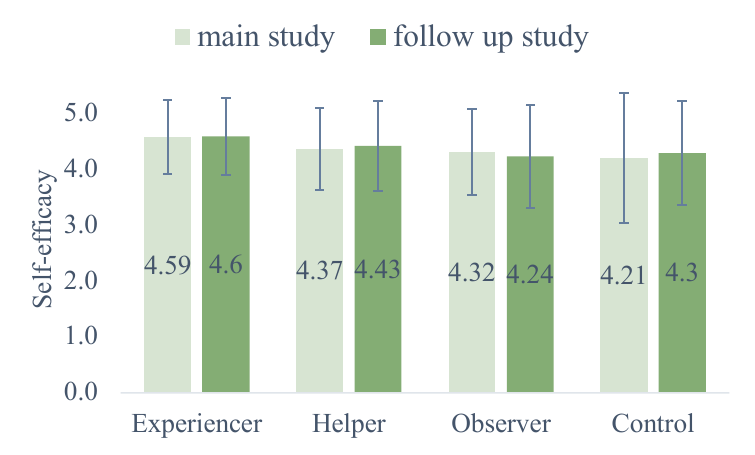}
        \caption{Self-efficacy (mean ± SD)}
        \label{SE_follow}
    \end{subfigure}
    
    \caption{Descriptive comparisons of effectiveness between the main study and the follow-up study.}
    \Description{Figure 7: Bar charts comparing the main study and the follow-up study. (a) Fraud cue identification scores (F1) remained stable across Experiencer, Helper, Observer, and Control groups, with little change between main and follow-up studies. (b) Self-efficacy scores also showed consistent patterns across groups between the two time points.}
    \label{main-follow}
\end{figure*}

    \modi{\autoref{main-follow} demonstrates that scores in all training conditions remained relatively stable from the main study to the follow-up, while the control group exhibited a modest, non-significant increase.} Moreover, for fraud cue identification, no significant within-group differences were found between the main and follow-up assessments across all groups, including Control (Z=1.660, \textit{P}=.102), Experiencer (Z=-0.284, \textit{P}=.798), Helper (t=-0.582, \textit{P}=.565), and Observer (t=0.331, \textit{P}=.745). Similarly, for self-efficacy, no significant changes were observed over time within any group, including Control (Z=-0.118, \textit{P}=.937), Experiencer (Z=-0.280, \textit{P}=.831), Helper (Z=0.560, \textit{P}=.582), and Observer (t=-0.347, \textit{P}=.733), shown in \autoref{within_group}.

\begin{table*}[htbp]
\centering
\caption{Within-group effectiveness differences between the main study and follow-up study.}
\Description{Table 8: Within-group comparisons of effectiveness between the main study and the follow-up study. For both fraud cue identification and self-efficacy, no significant differences were found within any group (Experiencer, Helper, Observer, Control) across the two time points.}
\begin{tabular}{llllll}
\toprule
 & Group       & Shapiro p-value & Test-used            & Test Statistic & P-value   \\
\midrule
\multirow{4}{*}{fraud cue identification} 
 & experiencer & 0.022           & Wilcoxon signed-rank & Z = -0.284     & .798  \\
 & helper      & 0.412           & Paired t-test        & t = -0.582      & .565    \\
 & observer    & 0.689           & Paired t-test        & t = 0.331      & .745    \\
 & control     & 0.023           & Wilcoxon signed-rank & Z = 1.660      & .102   \\
\midrule
\multirow{4}{*}{self-efficacy}            
 & experiencer & 0.027           & Wilcoxon signed-rank & Z = -0.280     & .831  \\
 & helper      & 0.010           & Wilcoxon signed-rank & Z = 0.560      & .582   \\
 & observer    & 0.102           & Paired t-test        & t = -0.347     & .733   \\
 & control     & 0.023           & Wilcoxon signed-rank & Z = -0.118     & .937  \\
\bottomrule
\end{tabular}
\label{within_group}
\end{table*}

    
    Mann-Whitney U tests did not find a significant difference for both fraud cue identification (W=489, \textit{P}=.169) and self-efficacy (W=602, \textit{P}=.863) between treatment groups and the control group. Similarly, Kruskal-Wallis tests indicated no significant difference, with specific results in \autoref{kw_effectiveness_follow}. In addition, there is no significant effectiveness difference between the three treatment groups shown in \autoref{kw_effectiveness_follow}.

    \textbf{Summary:} While the main study demonstrated significant group differences for fraud cue identification, \modi{these differences were not observed at the follow-up}. 
   The smaller sample sizes at follow-up may have reduced the statistical power, which could account for the absence of significant findings. \modi{Thus, we refrain from making strong claims about the underlying causes of the follow-up pattern.}
    
    

\begin{table}[htbp]
\centering
\caption{Results of the Kruskal–Wallis test comparing intervention effectiveness (follow-up study).}
\Description{Table 9: Results of Kruskal–Wallis tests on follow-up study effectiveness. No significant differences were found between treatment and control groups or among treatment groups for either fraud cue identification or self-efficacy.}
\begin{tabularx}{\columnwidth}{@{} p{1.5cm} p{3cm} c c @{}}
\toprule
 & Effectiveness Construct  & Test Statistic & P-value  \\
\midrule
\multirow{2}{*}{\makecell[l]{treatment\\ vs.\ control}} 
 & fraud cue identification & 2.95           & .399  \\
 & self-efficacy            & 2.07           & .558  \\
\midrule
\multirow{2}{*}{\makecell[l]{among\\ treatment}}       
 & fraud cue identification & 0.98           & .613  \\
 & self-efficacy            & 2.02           & .364  \\
\bottomrule
\end{tabularx}
\label{kw_effectiveness_follow}
\end{table}

\subsection{User Feedback of Intervention}

\subsubsection{Perceived Benefits of LLM-based Simulation}
\label{Perceived Benefits of LLM-based Simulation}
Participants expressed some positive views on using the LLM-based simulation for online fraud education. First, participants highlighted the \textbf{concreteness and detail} of the simulation, noting that conversations were intuitive, specific, and rich in information compared to traditional anti-fraud education. For example, P04 explained, 
\begin{quote}
    \textit{``This approach to fraud prevention feels more detailed, more targeted, and also more accurate, with a lot of concrete details. In all these details, you can see how to resist such behaviors. Today’s (LLM-based simulation) was quite detailed. When I give different content and prompts, its responses also vary. That way of chatting makes the experience sink in more deeply. It really sticks in the mind and leaves a stronger impression.''}
\end{quote}
P61 also remarked, \textit{``This is different from other methods that only provide fraud information. Here you actually chat with someone, which feels much better.''}

Second, the simulation was praised for its \textbf{flexibility and authenticity}. Participants described the interaction as flexible. As P140 noted,
\begin{quote}
    \textit{``Yours is definitely more flexible. Compared to traditional lectures, this one covers a much broader range. Lectures are really limited, narrow, and rigid, at least, very rigid. But with your system, I can let my thoughts wander, and it can also respond in unexpected ways, right? It’s more like that, because when you ask, it will always answer, isn’t that so? In that sense, the flexibility and liveliness are definitely much better.''}
\end{quote}
Some even perceived the dialogues as sincere and realistic. For example, P109 commented on its authenticity, \textit{``Because it (LLM-based simulation) seems very sincere and very real. Through the learning, we felt that the words generated by the AI sounded sincere. This kind of (learning) exchange can be done through AI.''}  


Third, participants emphasized the \textbf{practical relevance} of this learning approach, since fraudsters may also use virtual identities to interact with victims. P92 said, \textit{``I think it might be virtual. After all, scam groups always use something virtual to chat with people.''} Moreover, since the chances of real interactions with scammers were limited, they considered simulations an effective tool for fraud-prevention learning. The AI's ability to simulate scams was seen as ``just right'', not as overwhelming as a real scammer, but effective enough to serve as a reminder. P92 explained, 
\begin{quote}
    \textit{``For someone like me, who is hard to trick, your system didn't force me to keep chatting. With real scammers, the ones actually trying to get money, it would feel more exhausting. With the AI, I could just say a few words and then stop. I know very clearly that if something is a scam, I normally wouldn't believe it. So, I politely refused. The AI doesn't mind and won't keep chasing you, and you can still learn some knowledge about scams from it.''}
\end{quote}

Fourth, the simulation was considered \textbf{accessible for older adults}, especially those with lower educational backgrounds. P95 said, \textit{``This system is easier to understand. Even people with little education can follow.''}

Finally, several participants envisioned \textbf{future uses of AI in decision support}. They imagined shifting from the helper role, where older adults assist a potential victim, to AI directly supporting their own decision-making. For example, P109 suggested, \textit{``I think it (AI) is definitely more skilled than me in giving action advice. What if the AI persuades me instead? When facing suspicious situations, I could use AI to help me solve them.''}  P118 similarly hoped that if a friend were scammed, she could use AI to help persuade them more effectively, \textit{``If I realize my friends might be getting scammed and can’t convince you, I could bring in a robot to help persuade you. Because when you call the police, sometimes they can't give that many reasons. But a robot could be added to the conversation to help persuade your friend.''}

\subsubsection{Perceived Limitations of LLM-based Simulation.}
Despite these benefits, participants also pointed out several shortcomings of the LLM-based simulation. First, some described the interaction as \textbf{mechanical}, with responses that were too fast, overly official, or lacking natural flow. As P94 put it, \textit{``The replies were very official.''} P118 added, 
    \textit{``As soon as you say something, it immediately replies. There's still a bit of a mechanical feel in its responses, and the tone doesn't quite feel like a real person.''}

Second, participants noted a \textbf{lack of warmth} in the languages of LLM-based simulation. For example, P140 said, 
\begin{quote}
    \textit{``It's like a system, a programmed system, because it lacks the warmth that a real person has. You know, when scammers try to deceive, they put warmth into their language. So, I feel that in terms of language, that aspect is stronger than what you have here.''}
\end{quote}

Finally, some older adults also felt that the LLM showed \textbf{limited fraud sophistication}. For example, P92 said, \textit{``The kind of fake robot you mentioned, after all, it can't really compare with a real person. In terms of skills and techniques, it still falls a bit short.''} This limitation may also be related to our prompt design, since we explicitly restricted the simulation to predefined online fraud cues and did not allow the LLM to generate other cues, not only to prevent it from going off-topic, but also to avoid causing unforeseen harm to older adults.

\subsubsection{Experiencer Feedback}\label{Experiencer Feedback}

Participants found the Experiencer role valuable for immersive learning and enhancing self-protection confidence, though some also reported worries and mismatches with their personal preferences.

\textit{Positive feedback about experiencer.} \textbf{Immersive and realistic learning.} Several participants described the Experiencer role as a form of firsthand experience that allowed them to feel immersed in a simulated fraud situation. For example, P102 noted, \textit{``It was like we were actually experiencing it ourselves, like a scenario simulation. For instance, if you act as the scammer and you talk with me, we can really feel what that situation is like.''} P92 remarked, \textit{``It gave me the feeling of being personally involved.''} Participants also appreciated that this role helped them understand concretely how scammers operate. As P42 explained, \textit{``This included examples showing how scammers trick people, how they make up lies, and these were all there.''}

\textbf{Increased self-protection confidence.} Some experiencers reported that the role boosted their confidence in resisting scams. For example, P127 stated, \textit{``If someone really tried to scam me in that way in the future, they wouldn't succeed''}. P18 added, \textit{``Now that I've learned from your program and improved my knowledge. If I encounter such situations later, I won't be deceived.''}

\textit{Negative feedback about experiencer.} 
\textbf{Perceived worries.} Because the Experiencer role involved being ``scammed'' by a simulated fraudster, some participants reported feelings of worry. P127 directly explained, \textit{``He was trying to scam me and I felt unhappy.''} Others were concerned that they might actually fall for the scam and did not want to learn in this way. As P94 put it, 
\begin{quote}
    \textit{``Why go and learn from them? All they do is set traps for you. The more you chat with them, the more they start to use tricks. The deeper they go with their tricks, the more you keep chatting, and in the end, you'll naturally start to believe them and get carried away.''}
\end{quote} P18 said, \textit{``They are scammers, deceivers. Asking me to chat with them, but I don't even want to look at it.''} However, when we emphasized the educational purpose again, the same participant (P18) acknowledged, \textit{``If it is about reminders and warnings, I can fully accept it.''}

\textbf{Preference mismatch.} A few participants indicated that their personality made them dislike learning through communications. For example, P33 shared, \textit{``I don’t like this mode. My personality is a bit solitary.''} P115 said, \textit{``I simply don't like chatting.''}  

\subsubsection{Helper Feedback}\label{Helper Feedback}
In reflecting on the Helper role, participants emphasized its meaning in contributing to others and reinforcing their own skills, yet some also described confusion, disappointment, or difficulties in persuading the simulated victim.

\textit{Positive feedback about helper.} 
\textbf{Altruism and social responsibility.} Many participants described the Helper role as giving them a sense of altruism and social contribution. They valued the opportunity to share their own experiences and help others, aligning this activity with broader cultural values of ``helping others'' in China. P131 explained, 
\begin{quote}
    \textit{``Helping others gave me a sense of pleasure and altruism. When it comes to preventing fraud, if older adults are being deceived, and you have good experience to share, why not provide it, as long as it's the right way? If you can reduce the chance of someone around you being scammed, why wouldn't you do it? You don't suffer any loss, right? And besides, in China we advocate for helping others with joy.''}
\end{quote} Some participants even suggested that after practicing in the helper role, they would be able to extend this behavior to real life, becoming advocates for anti-fraud education in their communities. P131 future shared \textit{``We start by persuading a virtual person, but once we learn it, we can also persuade real people.''} Similarly, P36 noted, \textit{``If a friend is being scammed, I could help stop them after learning. That is very practical.''}

\textbf{Sense of achievement.} Participants often reported feeling a sense of accomplishment when they successfully persuaded the simulated victim. For example, P134 said, \textit{``It gives me a sense of achievement, if I manage to persuade that person. I really feel a sense of accomplishment inside.''}

\textbf{Self-reinforcement.} Several participants reflected that helping others also reinforced their own awareness. For example, P141 noted, \textit{``By helping others, I also improve my own ability to recognize scams.''}

\textit{Negative feedback about helper.}
\textbf{Role confusion and skepticism.} Some participants were unsure about the identity of the simulated victim, worrying that the victim might itself be deceptive. P140 explained, \textit{``If you want to ask, you should consult someone knowledgeable, right? You shouldn't come to me, trying to drag me in. My first response to him was: I don't know, I won't get involved.''} Even after we repeatedly emphasized that they were playing the helper role and the other party was a victim, some still insisted on interpreting them as scammers. P129 insisted, \textit{``But I still don't have the ability to recognize it. To me, he just seemed like a scammer. Talking about high income \dots It’s all fake. Those kinds of financial investments are high risk. I felt it was more like he was trying to scam the next person.''}

\textbf{The sense of disappointment.} While some participants said that persuading others gave them a sense of achievement, others mentioned that failing to persuade created a sense of disappointment. For example, P131 said, \textit{``If you succeed in persuading someone, of course you feel a sense of achievement. But if you can't persuade them, then you don't feel any achievement at all. That kind of gap, that letdown from not achieving it, feels quite big.''}

\textbf{Difficult to persuade.} A complaint was that the simulated victim was hard to persuade. For example, P111 shared, 
\begin{quote}
    \textit{``Since you asked me to persuade him, I just told him it was a scam, a fraud. But let's say he really, really believed in it, and he kept circling around that topic and wouldn't let go. With people, after you say a couple of sentences like `this is a scam' a normal person might stop and think, `Oh, maybe I should reconsider.' But here, it felt like he was too deeply poisoned. It's not the same as chatting with a real person. Because with the system, one question will just keep coming back again and again.''}
\end{quote} This effect may stem from our prompt design, which explicitly instructed the LLM to elicit persuasion attempts from older adults.

\subsubsection{Observer Feedback}\label{Observer Feedback}

Reactions to the Observer role varied, with many participants highlighting its usefulness for awareness and reflection, while others considered it less engaging.

\textit{Positive feedback about observer.} 
\textbf{Warning effect.} Participants noted that observing others being scammed had a strong warning and educational effect. It reinforced the impression that older adults like themselves could easily become victims. P133 explained, \textit{``Just by watching, the impact was quite strong because it had an educational meaning. We realized we could also easily fall for it.''} P138 added, \textit{``These cases have a lot of warning value, and I think they still achieve a certain effect.''}  

\textbf{Empathy and reflection.} Several older adults reported feeling empathy when they saw others being scammed, which in turn prompted them to reflect and learn lessons for themselves. P138 said, \textit{``When you see someone get scammed, you feel empathetic. Because you also need to guard against being scammed yourself.''} P49 emphasized the value of reflection, \textit{``Seeing this, we all learn a lesson from it.''} P116 added, 
\begin{quote}
    \textit{``Others were scammed. You need to reflect and summarize, right? Sometimes if you don't reflect, you won't notice, and you'll get scammed yourself. You have to draw lessons from others' experiences. Honestly, no matter how smart you are, if you’re not paying attention, you can still be scammed.''}
\end{quote}

\textit{Negative feedback about observer.}  
\textbf{Low engagement.} Some participants found the Observer role less engaging or even boring. As P70 put it, \textit{``It was not interesting to learn this way. I didn’t like it.''}

\section{Discussion}

\subsection{Reimagining Fraud Education for Older Adults through Role-based Simulation}
Compared with traditional anti-fraud education that relies on passive information delivery, role-based simulation offers hands-on and first-person perspectives, realistic contexts, and greater engagement \cite{lateef2010simulation,shaw2018designing,url6}. Although prior work has shown the effectiveness of role-based simulation for younger or general populations \cite{chen2024effects, wen2019hack, kumaraguru2008lessons}, its applicability to older adults has remained unclear, especially given that they may be less inclined toward interactive ways of S\&P education \cite{aly2024tailoring}. Our findings demonstrate that role-based simulation is both feasible and effective for older adults in improving fraud cue identification relative to a control group (Sec. \ref{Online Fraud Cue Identification Comparison}). The results suggest that older adults may be more willing and able to benefit from interactive approaches than previously assumed, highlighting role-based simulation as a promising direction for future fraud interventions for older adults, offering a more engaging alternative to passive knowledge consumption.

\subsubsection{Moving Beyond General Information: Fine-grained Fraud Education}
In terms of simulation content, prior work on fraud education for older adults has largely focused on relatively general fraud information
\cite{chung2023reducing,camilleri2025cybersafety,url4,url5}. While informative, such general knowledge does not necessarily equip older adults with the ability to recognize the detailed cues that often signal fraud, which may hinder their ability to detect scams in practice. Role-based simulation offers a distinct advantage by exposing learners to the fine-grained details of scam interactions, enabling them to develop practical recognition skills in realistic fraud scenarios. In this study, we organized and extracted linguistic fraud cues to embed into the simulations. Beyond language, however, scams often involve multimodal details, such as deepfakes \cite{zhai2025hear}. Incorporating such fine-grained and multimodal cues into role-based simulation could give older adults a richer and more realistic understanding of how online fraud operates. In addition, our current simulations relied on a fixed set of curated fraud cues. A promising direction would be to leverage AI agent capabilities to dynamically update fraud cues based on emerging scam tactics and to adapt training to individual participants' weaknesses.

\subsubsection{Leveraging LLMs for Role-based Simulation: Opportunities and Challenges}
Role-based simulation can be effectively implemented through large language models. 
Older adults found several perceived benefits, describing the simulation as concrete and detailed, the conversations as intuitive, rich, and realistic, and the overall dialogue format as flexible and easy to understand (Sec. \ref{Perceived Benefits of LLM-based Simulation}). Overall, LLMs proved capable of playing simulation roles effectively and were generally well received by older adults. Looking ahead, role-based simulation powered by LLMs appears to be a promising approach for online fraud education among older adults. At the same time, participants also noted limitations. For example, some perceived the dialogues as somewhat mechanical or lacking emotional warmth, suggesting the possibility of incorporating affective elements to create more authentic simulations~\cite{fang2025social}. However, future work should carefully consider how to balance simulation authenticity with instructional effectiveness - the ultimate goal is not to create a perfectly realistic simulation, but rather to develop an experience that maximizes learning outcomes while maintaining sufficient realism to engage users. 

Special caution is needed when applying LLM-based simulations, as the model may occasionally produce unexpected outputs \cite{rapp2025people} that could potentially cause harm to older adults. In our study, this risk was managed by having researchers present throughout the intervention to immediately stop the session if any inappropriate content appeared. While this safeguard proved effective, it may not be feasible for large-scale deployment. Future implementations will therefore need to explore additional protective measures, such as integrating a monitoring model to review the primary model's outputs in real time and automatically interrupt or replace unsafe responses, or adopting a human-in-the-loop approach in which community staff, family members, or other facilitators are present to ensure safe use.

 \subsubsection{Generalizability}

\modi{Our fraud cues are derived from real scam cases and grounded in the specific vulnerabilities of older adults. For example, the cue ``high returns with low risk'' corresponds to the vulnerability related to financial expectations. Many different fraud types are, in fact, built upon the same underlying vulnerabilities \cite{deng2025auntie}, which means that the cues we identified can be, to some extent, generalized to educational interventions targeting other fraud types that exploit financial expectations or health and well-being concerns in real-world scenarios. More broadly, our work offers a transferable approach that extracts fraud cues through the lens of user vulnerabilities and uses these cues as the basis for fraud education. This approach could be adopted by future educators or researchers who aim to design fraud-prevention interventions, especially in chat-based, step-by-step manipulative scam contexts. They could apply the same logic to identify fraud cues associated with the specific vulnerability they intend to target, and then use those cues to support tailored educational goals.} 

Nonetheless, because the fraud cues and learning materials in our study were based on the Chinese context, the extent to which role-based simulation for online fraud education can be generalized to other cultural settings remains an open question. Future work could build on fraud cues and prevalent fraud contexts in other regions to examine the effectiveness of such interventions across cultures. Beyond cultural adaptation, generalizability to other populations is also worth exploring. Prior research has suggested that younger populations tend to be receptive to chatbot-based and interactive learning \cite{aly2024tailoring}, and simulated phishing has already been studied with general populations \cite{nasser2020role,downs2006decision,sheng2007anti}. Thus, LLM-based role-based simulation may also hold promise beyond older adults.
 
\subsection{Broadening Fraud Education through Multi-Role Perspectives}
Most prior role-based simulations in fraud education have predominantly focused on the victim role, aiming to help participants learn by experiencing fraud directly \cite{karumbaiah2016phishing,lain2022phishing,gopavaram2021cross}. However, learning is not restricted to those who undergo the experience themselves. For example, prior work in healthcare simulation has shown that observer roles can achieve learning outcomes that are as effective or even superior to those in hands-on roles \cite{o2016observer}. Similarly, beyond the experiencer, the helper role also proves valuable as helpers outperform the control group in fraud cue identification (see Sec. \ref{Online Fraud Cue Identification Comparison}). This multi-perspective approach to learning may provide users with distinct experiences and insights.

\subsubsection{Experiencer}
Older adults reported that the experiencer role gave them an immersive, first-hand sense (Sec. \ref{Experiencer Feedback}). Quantitatively, this role was also associated with higher effectiveness than the control group in fraud cue identification (Sec. \ref{Online Fraud Cue Identification Comparison}). This finding resonates with experiential learning theory, which emphasizes ``learning by doing'' through a hands-on approach \cite{kolb2014experiential}. Prior research has similarly demonstrated the effectiveness of experiential approaches in domains such as dart patterns \cite{ye2025awareness}, and our results extend this evidence to the context of fraud education. We therefore suggest that future cybersecurity education should integrate experiential components to foster deeper learner understanding. However, special consideration is needed when applying this role with older adults. Some participants expressed concerns about ``actually being scammed'' during the exercise (Sec. \ref{Experiencer Feedback}). Once the learning purpose was clarified, they became more willing to accept this role. Therefore, future implementations should provide older adults with clearer and more salient explanations, emphasizing that the primary goal is to help them recognize how scams may unfold.

\subsubsection{Helper}
We found that teaching others enabled helpers to better identify fraud cues (Sec. \ref{Online Fraud Cue Identification Comparison}). This aligns with prior research demonstrating that peer teaching strengthens the teacher's own understanding of the material \cite{galbraith2011peer,hoogerheide2016gaining}. Compared to simply being taught, the act of explaining requires individuals to actively review and reformulate knowledge, similar to the way people consolidate their own thoughts by explaining them to friends \cite{duran2017learning,debbane2023learning}. 
Moreover, we observed that the helper role was enriched by feelings of altruism and social responsibility (Sec. \ref{Helper Feedback}). Many older adults reported a sense of achievement after helping, echoing prior findings that older adults tend to exhibit greater altruism than younger adults \cite{sparrow2021aging}. This prosocial orientation may amplify the effectiveness of the helper role among older adults, which also sheds light on the potential of helper role in broader simulation-based learning.

Despite the benefits of the helper role, several challenges deserve attention in future work. Some helpers expressed doubts about their role, including concerns that the person they were assisting might actually be a scammer attempting to entrap them (Sec. \ref{Helper Feedback}). This highlights the importance of clearly communicating to helpers that the other party is a victim in need of assistance, rather than a potential threat. One potential solution is to particularly note differences between task-related interactions and prior background information. Besides, some helpers reported disappointment when their persuasion efforts were unsuccessful (Sec. \ref{Helper Feedback}). Future work should pay careful attention to the emotional burden that the helper role may place on older adults, particularly when they encounter scenarios involving distressed victims or feel responsible for others' safety. In order to preserve older adults' confidence, future iterations may benefit from a gamified design, where persuasion tasks are structured with varying levels of difficulty to ensure that participants experience achievable successes while still being challenged.

\subsubsection{Observer}

Older adults reported that simply observing produced a warning effect and enabled empathy and self-reflection (Sec. \ref{Observer Feedback}). This aligns with the principles of social learning theory, which posits that observation alone without direct practice can still facilitate learning \cite{bandura1977social}. However, our quantitative results did not show that observers outperformed participants in other conditions. One possible reason is that, without interactive engagement, older adults may only gain a general sense of the risks of online fraud while struggling to pinpoint specific cues of danger. Prior work suggests that observation can be more effective when supported by appropriate observing tools \cite{o2016observer}. Building on this, future interventions could provide observers with lightweight scaffolds, for instance, a visual annotation tool or mini-quizzes alongside the interaction, that might help them more actively process and retain what they observe.

Overall, the different perspectives afforded by distinct roles bring unique benefits as well as challenges. Future work in fraud education could explore such multi-perspective approaches, as they may help learners step outside of their potential victim roles and view the same problem from alternative standpoints, fostering different insights and deeper reflection. We thus advocate that future fraud education initiatives incorporate a broader range of viewpoints to enrich learning outcomes.

\subsection{Implications}

\subsubsection{Design Implications}

\textit{Integrating different roles into one application.} Preferences for different roles may not only stem from the roles themselves but also from individual characteristics such as personality. For example, we observed that some older adults in the experiencer group reported that they were more introverted and therefore did not enjoy chat-based interactions (Sec. \ref{Experiencer Feedback}). For them, the observer role might have been more suitable. Since our study followed a between-group design, each participant engaged with only one role, and thus we could not directly compare within-person preferences across roles. Future work could explore integrating multiple roles into a single educational application, allowing older adults to freely select and experiment with different roles. Such flexibility could help individuals discover one or more formats that best align with their preferences and approach the same problem from multiple perspectives, supporting more well-rounded learning.



\textit{Mitigating confusion between simulation and reality.} Both in Experiencer and Helper modes, a few older adults expressed concerns about being scammed for real. While researchers provided immediate clarification during our study, this highlights a critical design challenge for simulation-based fraud education: balancing realistic engagement with clear educational framing. Future work should carefully consider the balance between maintaining sufficient detachment (to ensure users understand they are in a simulation) and preserving immersion (to maintain engagement and learning effectiveness). For example, the system could use distinct visual cues (such as educational logos) to constantly remind users of the simulation environment without disrupting the experience. Additionally, when deploying such systems without human supervision, careful attention must be paid to the onboarding process. This should include explicit educational framing, clear explanations of the simulation's purpose and boundaries, step-by-step previews of the interaction format, and require participants to confirm their understanding of the educational nature of the exercise.

\textit{Scaling role-based interventions.} Future work could explore the scalability of the current intervention. Our experimental design focused on two scenarios, health and finance, but many other fraud contexts exploit older adults' vulnerabilities, such as overreliance on authority, social isolation, and generational affection (Sec. \ref{fraud cue}). These scenarios could be readily incorporated into the intervention by organizing additional fraud cues. Moreover, the scalability could also extend to the simulation channels. In our current design, WeChat served as the messaging interface for role-based simulation. However, 
livestreaming platforms represent another major site of online fraud targeting older adults. Such platforms could similarly support multi-role perspectives, for instance, by simulating audiences or content moderators as additional roles. Finally, scalability might also apply to the interaction patterns within each role. For example, in the helper condition, ``learning by teaching'' can take multiple forms, including teaching one peer, teaching a group of peers, or teaching collaboratively in a group where each member becomes an expert on a different topic \cite{slavin1980cooperative,debbane2023learning}. Future research could explore whether adapting these different forms, such as helping one victim, assisting a group, or collaborating with other older adults to provide help, might further improve learning outcomes in online fraud education.

\subsubsection{Educational Implications}
\textit{Embedding role-based simulation into existing fraud education.} From a practical perspective, our intervention may offer a complementary pathway to support existing anti-fraud education for older adults. Governments, communities, and NGOs have already developed extensive resources, such as brochures, training sessions, and public lectures, that provide valuable strategies and examples. These resources, however, are often presented in traditional formats and may not always encourage active engagement. Our role-based simulation approach could serve as a complementary layer, enabling existing educational content to be experienced in a more interactive and multi-perspective manner. Rather than replacing current practices, this approach might be explored as a way of enriching them, with the potential to make fraud education for older adults more engaging and impactful in real-world contexts.

\textit{Advocating for the longitudinal effects.} Role-based simulation learning proved more effective than the control condition after the intervention (Sec. \ref{Online Fraud Cue Identification Comparison}). \modi{However, the follow-up evaluation did not reproduce the significant group differences found in the main study (Sec. \ref{Short-effects of Intervention}).
Although the follow-up results diverged from the initial pattern, this finding underscores the importance of incorporating longitudinal evaluation when studying online fraud education. Cross-sectional post-intervention gains may not fully reflect the stability or durability of learning outcomes, and without follow-up assessments, potential fluctuations, whether due to sampling constraints or other contextual factors, may go unnoticed.} Our reflection extends to prior educational intervention studies, where evaluations of longitudinal effects are often lacking. Therefore, we advocate for future research to systematically incorporate longitudinal evaluation in order to capture the longitudinal effects of learning outcomes and better inform the design of durable and effective intervention strategies.

\subsection{Limitations}
First, all fraud cues and learning materials were derived from Chinese contexts, and the participants were also Chinese. Therefore, the findings may not be generalizable and require further validation in other cultures. Second, the non-significant difference in user experience scores may be due to the observation that older adults were very appreciative of our fraud-prevention efforts, and thus reported positive experiences regardless of the learning modes. Third, self-reported user experience might have the potential social-desirability bias \cite{dell2012yours}, so future work can explore observation techniques \cite{distler2023influence} and use more objective metrics (e.g., the time spent interacting with the interface) to compare different learning modes. Fourth, we only compared the role-based simulation with a control group (reading educational material as training); future work could examine the effects of role-based simulation versus in-person workshops \cite{chen2024effects} or educational video series \cite{volkamer2018developing}, as these have been evaluated in prior work as effective cybersecurity training approaches for other demographic groups. \modi{Finally, we include only two representative contexts (finance and health) and call for further validation across other contexts.}

\label{limitations}

\section{Conclusion}
In this work, we designed three perspectives of role-based simulation for older adults' fraud education. In a between-subjects experiment with 144 participants, Experiencer and Helper roles effectively improved fraud cue identification, with Helper outperforming Observer. \modi{However, the follow-up results did not replicate the initial group effects, highlighting the value of incorporating longitudinal evaluation when designing interventions.} From older adults' perceptions, LLM-based multi-role simulation is feasible but also presents challenges, and different roles offer distinct benefits and drawbacks. We discuss future directions for fraud education through role-based simulation, highlighting multi-role perspectives and implications for design and practice.

\begin{acks}
The research is partially funded by the Deutsche Forschungsgemeinschaft (DFG, German Research Foundation) under Germany’s Excellence Strategy - EXC 2092 CASA - 390781972, Google Academic Research Award on Trust and Safety (2024; ID 00029925), NSFC grant 62432008, RGC RIF grant R6021-20, RGC TRS grant T43-513/23N-2, RGC CRF grant C6015-23G, NSFC/RGC grant CRS\_HKUST601/24, and RGC GRF grants 16207922, 16207423 and 16203824. We thank Gabriel Lima and Tarini Saka for their valuable feedback on our earlier drafts. 
\end{acks}



\bibliographystyle{ACM-Reference-Format}
\bibliography{sample-base}

\appendix
\newpage
\section{Fraud Cue Identification Test}
\label{fraud cue identification test}
\subsection{Fraud Cue Identification Test (Main Study)}
\label{Cases in Fraud Awareness Assessment (Main Study)}

\textbf{\modi{Finance A Test}}
\begin{figure}[H]
	\centering
		\includegraphics[width=0.5\textwidth]{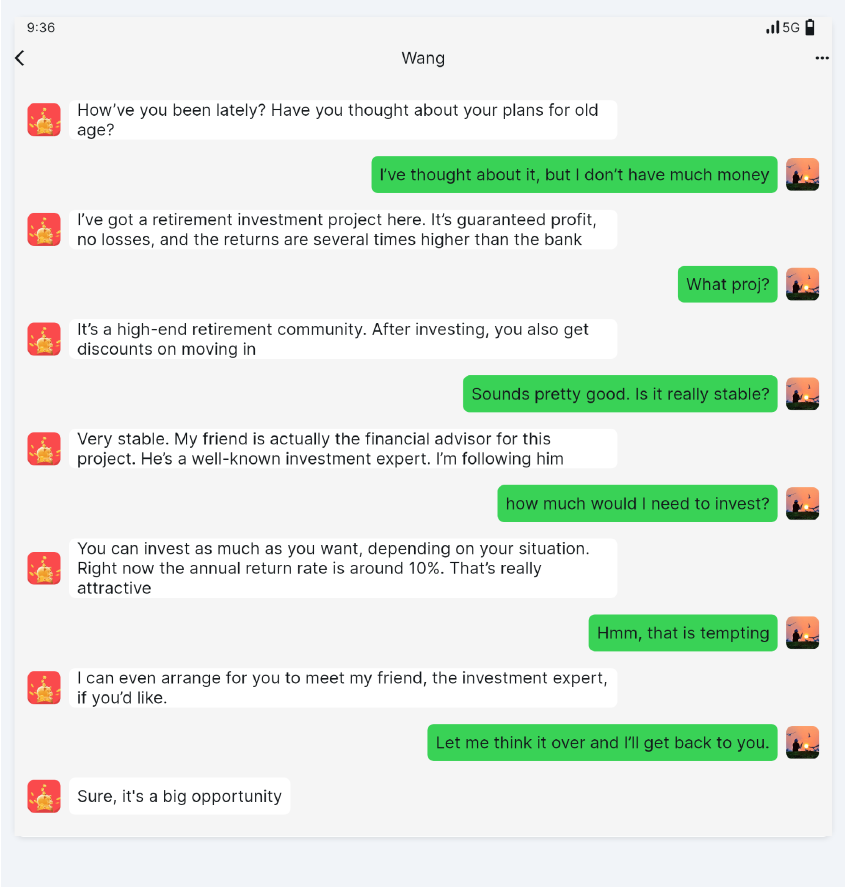}
	\caption{Finance A Test}
    \Description{Figure 8: A specific test case in the main study for finance A.}
\end{figure}

\modi{\textbf{Q: Please select the fraud cues below that you think are present in the above conversation}
\begin{itemize}
    \item \textbf{Promise high returns with low risk}
    \item \textbf{Claim to be financial experts}
    \item Leverage peer influence
    \item \textbf{Pitch senior-living investment schemes}
    \item Invoke fake government investment schemes
    \item Promise dividends between shareholders
    \item None of the above
\end{itemize}}

\newpage
\textbf{\modi{Finance B Test}}
\begin{figure}[H]
	\centering
		\includegraphics[width=0.5\textwidth]{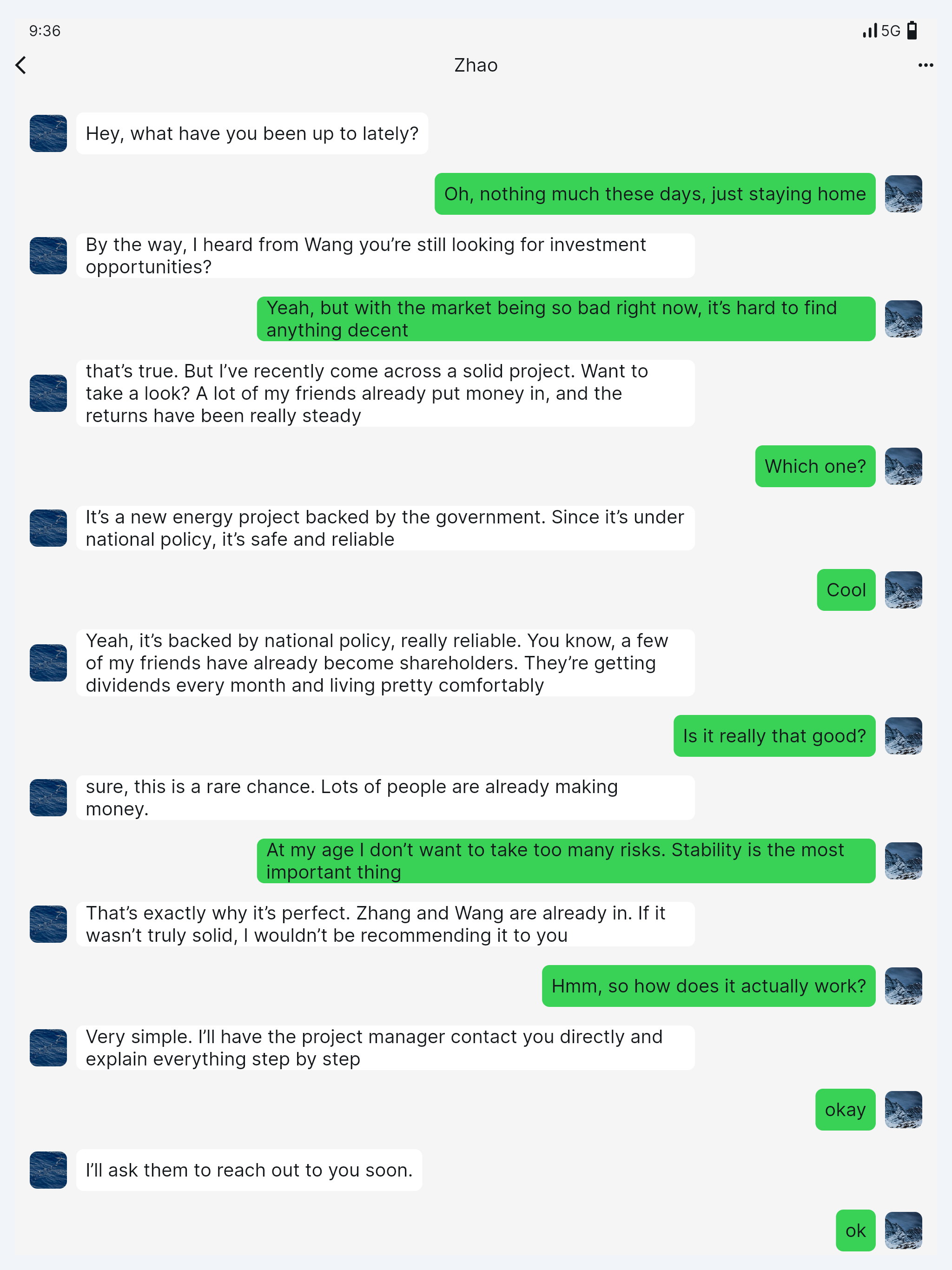}
	\caption{Finance B Test}
    \Description{Figure 9: A specific test case in the main study for finance B.}
\end{figure}

\modi{\textbf{Q: Please select the fraud cues below that you think are present in the above conversation}
\begin{itemize}
    \item Promise high returns with low risk
    \item Claim to be financial experts
    \item \textbf{Leverage peer influence}
    \item Pitch senior-living investment schemes
    \item \textbf{Invoke fake government investment schemes}
    \item \textbf{Promise dividends between shareholders}
    \item None of the above
\end{itemize}}

\newpage
\textbf{\modi{Health A Test}}
\begin{figure}[H]
	\centering
		\includegraphics[width=0.5\textwidth]{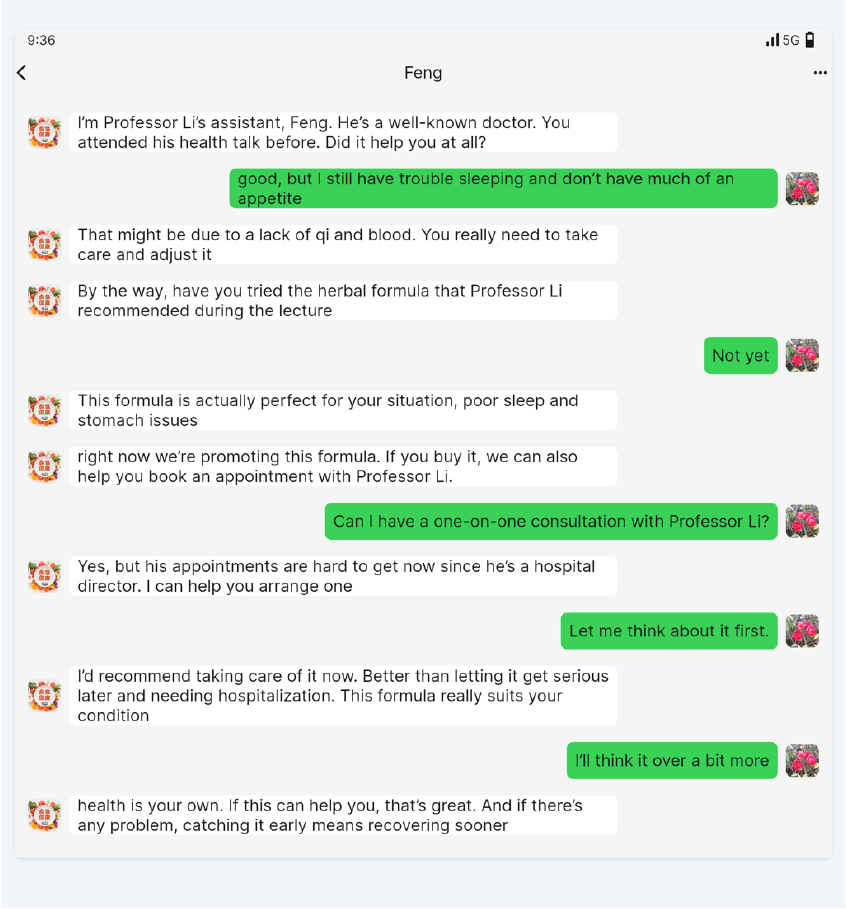}
	\caption{Health A Test}
    \Description{Figure 10: A specific test case in the main study for health A.}
\end{figure}

\modi{\textbf{Q: Please select the fraud cues below that you think are present in the above conversation}
\begin{itemize}
    \item \textbf{Exaggerate illness to create anxiety}
    \item \textbf{Advertise referrals to top doctors}
    \item Offer free services as bait
    \item Using cheap trials with steep follow-up charges
    \item \textbf{Claim to be medical experts}
    \item Frame products as miracle cures with refund guarantees
    \item None of the above
\end{itemize}}

\newpage
\textbf{\modi{Health B Test}}
\begin{figure}[H]
	\centering
		\includegraphics[width=0.5\textwidth]{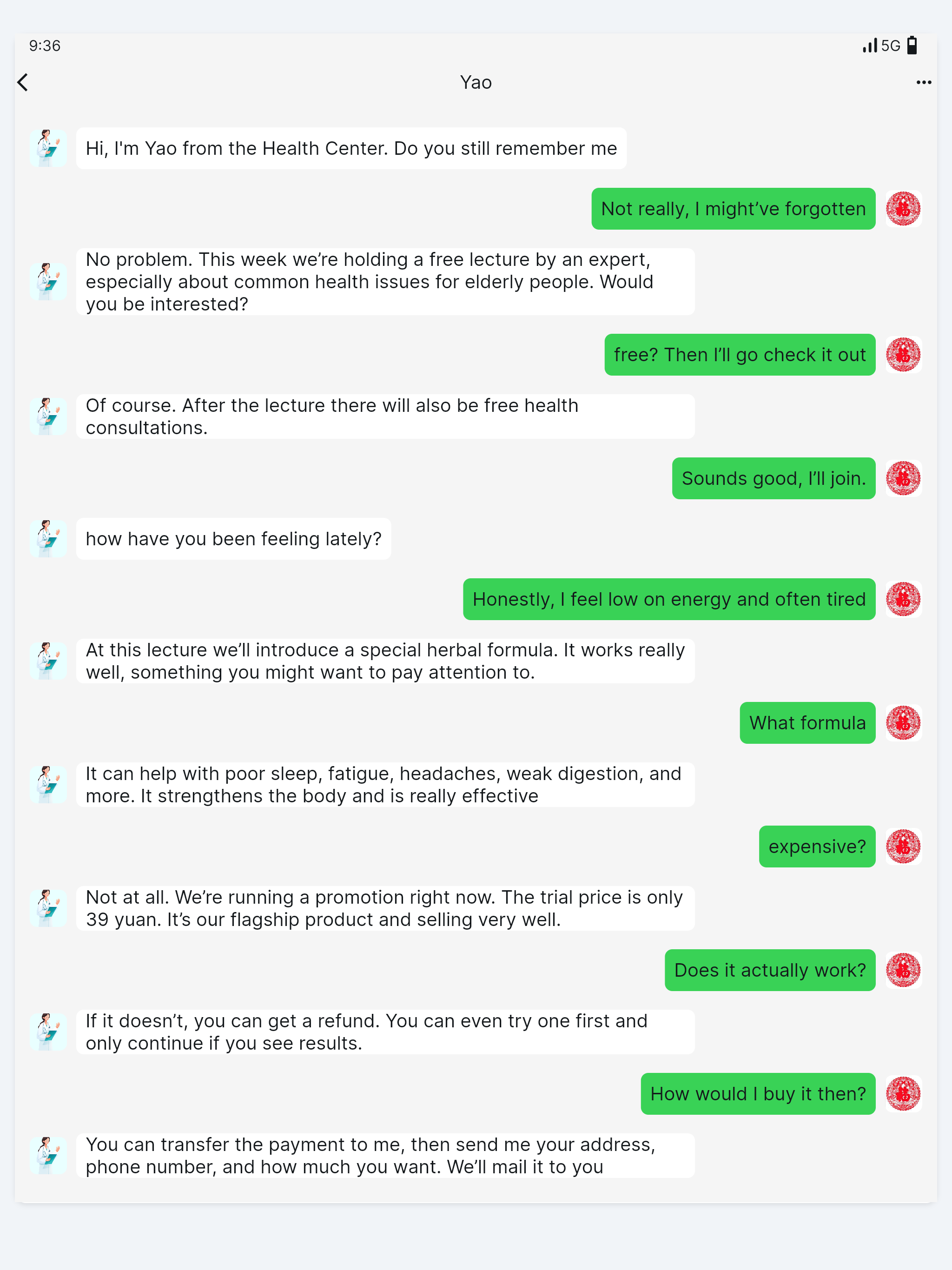}
	\caption{Health B Test}
    \Description{Figure 11: A specific test case in the main study for health B.}
\end{figure}

\modi{\textbf{Q: Please select the fraud cues below that you think are present in the above conversation}
\begin{itemize}
    \item Exaggerate illness to create anxiety
    \item Advertise referrals to top doctors
    \item \textbf{Offer free services as bait}
    \item \textbf{Using cheap trials with steep follow-up charges}
    \item Claim to be medical experts
    \item \textbf{Frame products as miracle cures with refund guarantees}
    \item None of the above
\end{itemize}}

\newpage
\subsection{Fraud Cue Identification Test (Follow-up Study)}
\label{Cases in Fraud Awareness Assessment (Follow-up Study)}

\textbf{\modi{Finance A Test}}
\begin{figure}[H]
	\centering
		\includegraphics[width=0.5\textwidth]{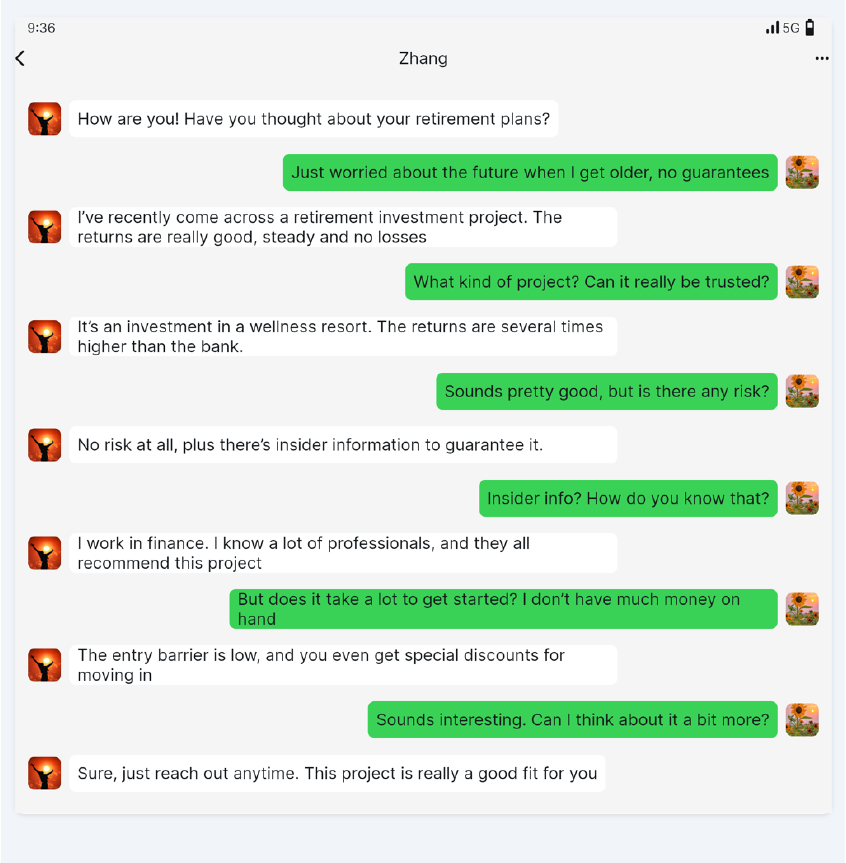}
	\caption{Finance A Test}
    \Description{Figure 12: A specific test case in the follow-up study for finance A.}
\end{figure}

\modi{\textbf{Q: Please select the fraud cues below that you think are present in the above conversation}
\begin{itemize}
    \item \textbf{Promise high returns with low risk}
    \item \textbf{Claim to be financial experts}
    \item Leverage peer influence
    \item \textbf{Pitch senior-living investment schemes}
    \item Invoke fake government investment schemes
    \item Promise dividends between shareholders
    \item None of the above
\end{itemize}}
\newpage
\textbf{\modi{Finance B Test}}
\begin{figure}[H]
	\centering
		\includegraphics[width=0.5\textwidth]{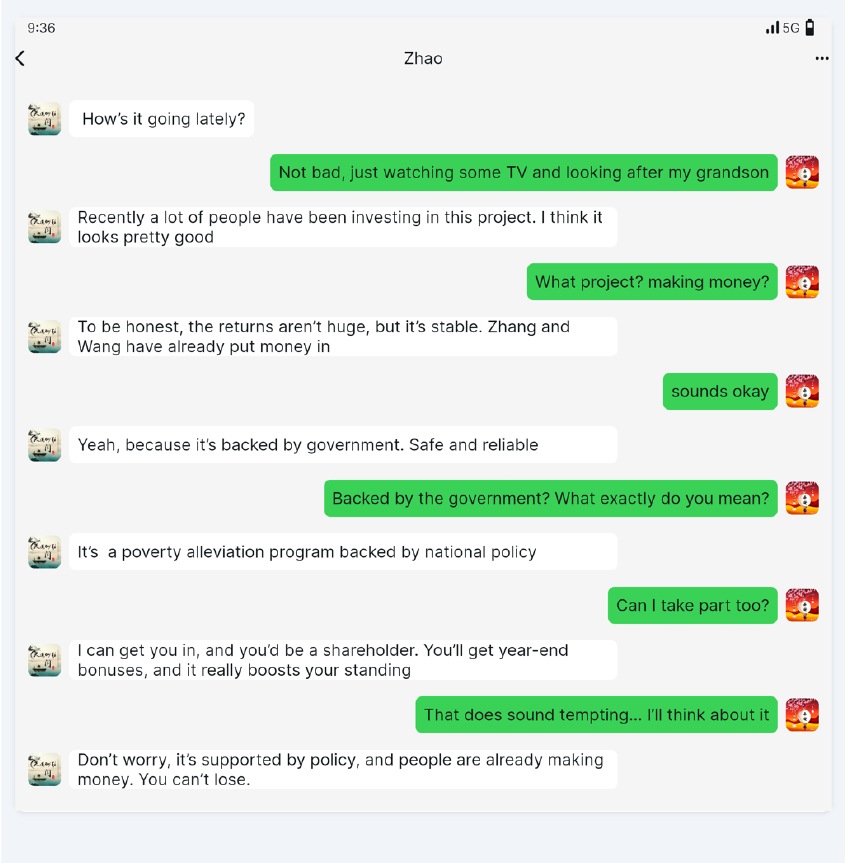}
	\caption{Finance B Test}
    \Description{Figure 13: A specific test case in the follow-up study for finance B.}
\end{figure}

\modi{\textbf{Q: Please select the fraud cues below that you think are present in the above conversation}
\begin{itemize}
    \item Promise high returns with low risk
    \item Claim to be financial experts
    \item \textbf{Leverage peer influence}
    \item Pitch senior-living investment schemes
    \item \textbf{Invoke fake government investment schemes}
    \item \textbf{Promise dividends between shareholders}
    \item None of the above
\end{itemize}}

\newpage
\textbf{\modi{Health A Test}}
\begin{figure}[H]
	\centering
		\includegraphics[width=0.5\textwidth]{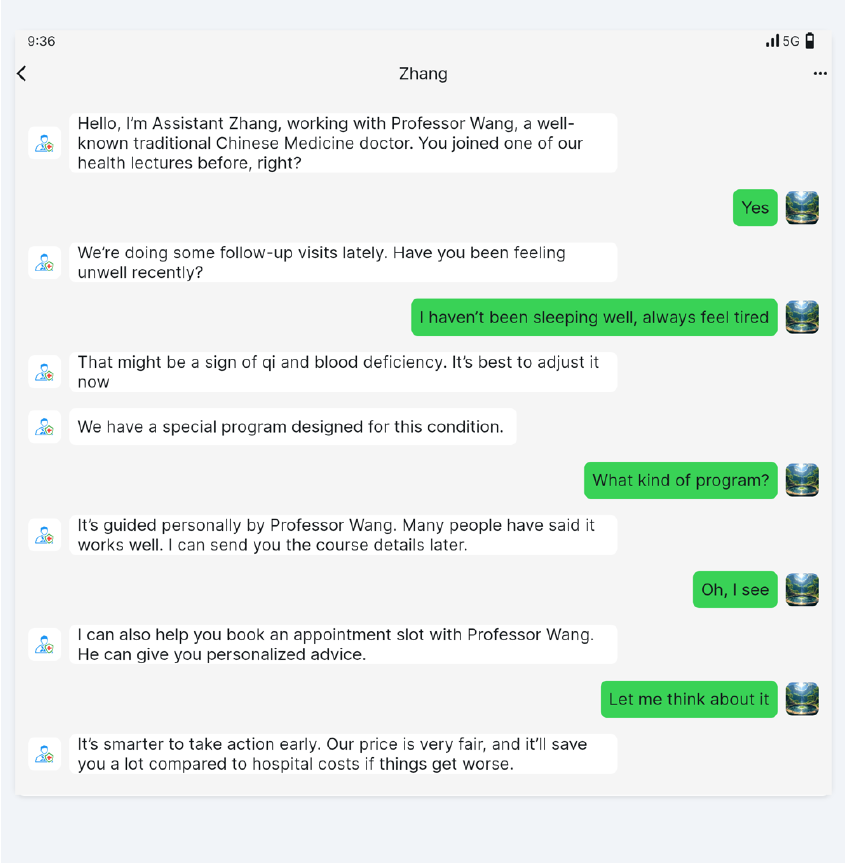}
	\caption{Health A Test}
    \Description{Figure 14: A specific test case in the follow-up study for health A.}
\end{figure}

\modi{\textbf{Q: Please select the fraud cues below that you think are present in the above conversation}
\begin{itemize}
    \item \textbf{Exaggerate illness to create anxiety}
    \item \textbf{Advertise referrals to top doctors}
    \item Offer free services as bait
    \item Using cheap trials with steep follow-up charges
    \item \textbf{Claim to be medical experts}
    \item Frame products as miracle cures with refund guarantees
    \item None of the above
\end{itemize}}

\newpage
\textbf{\modi{Health B Test}}
\begin{figure}[H]
	\centering
		\includegraphics[width=0.5\textwidth]{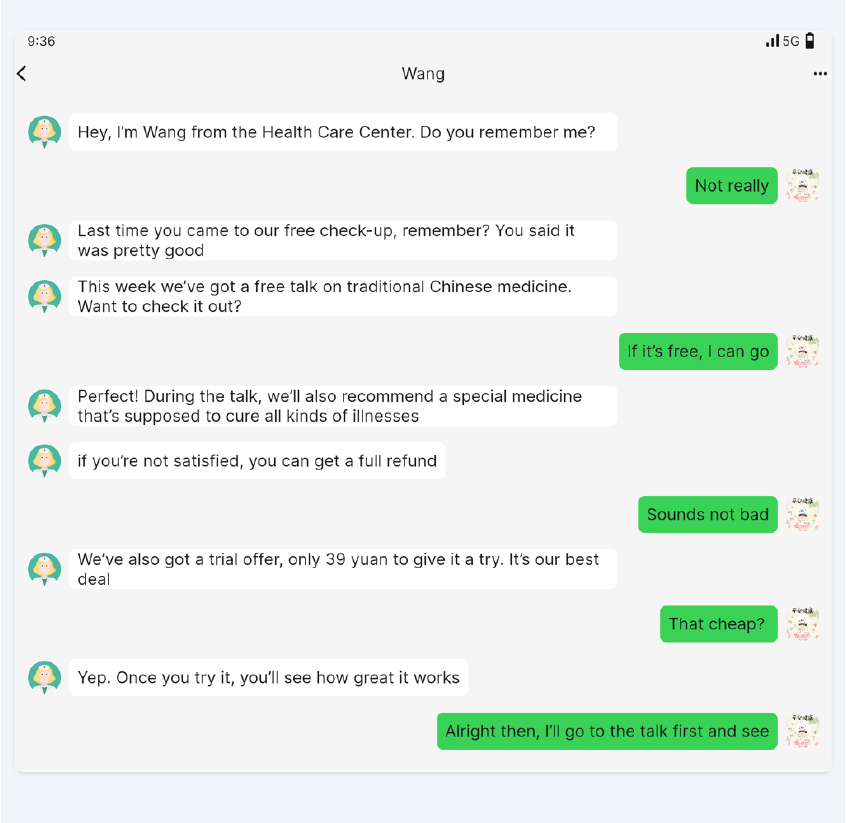}
	\caption{Health B Test}
    \Description{Figure 15: A specific test case in the follow-up study for health B.}
\end{figure}

\modi{\textbf{Q: Please select the fraud cues below that you think are present in the above conversation}
\begin{itemize}
    \item Exaggerate illness to create anxiety
    \item Advertise referrals to top doctors
    \item \textbf{Offer free services as bait}
    \item \textbf{Using cheap trials with steep follow-up charges}
    \item Claim to be medical experts
    \item \textbf{Frame products as miracle cures with refund guarantees}
    \item None of the above
\end{itemize}}

\newpage
\section{Scale Questions in Evaluation}
\label{scale question}
\begin{table}[H]
\caption{Measurement items used in the evaluation.}
\Description{Table 10: Measurement items used in the evaluation. Constructs include self-efficacy, enjoyment, usability, and future behavioral intention. Each construct is assessed with multiple items.}
\centering
\small
\setlength{\tabcolsep}{6pt}
\begin{tabularx}{\columnwidth}{@{} p{1.7cm} p{6.5cm}@{}}
\toprule
\textbf{Construct} & \textbf{Items} \\
\midrule
\multirow{2}{*}{self-efficacy \cite{ormond2016perceived}} 
  & I am confident I can detect online scams. \\
  & I am able to detect online scams without much effort. \\
\midrule
\multirow{4}{*}{enjoyment \cite{agarwal2000time}} 
  & I have fun interacting with the system. \\
  & Using the system provides me with a lot of enjoyment. \\
  & I enjoy using the system. \\
  & Using the system bores me. \\

\midrule
\multirow{4}{*}{usability \cite{finstad2010usability}} 
  & This learning system’s capabilities meet my requirements. \\
  & Using this learning system is a frustrating experience. \\
  & This learning system is easy to use. \\
  & I have to spend too much time correcting things with this learning system. \\
\midrule
\multirow{3}{*}{\makecell[l]{future behavioral \\intention \cite{agarwal2000time}}} 
  & I plan to use the system in the future. \\
  & I intend to continue using the system in the future. \\
  & I expect my use of the system to continue in the future. \\
\bottomrule
\end{tabularx}
\end{table}

\end{document}